\documentclass[aps, pre, twocolumn, superscriptaddress]{revtex4-2}
\usepackage{graphicx}
\usepackage{amsfonts}
\usepackage{amsmath}
\usepackage{amssymb}
\usepackage{hyperref}
\usepackage{xcolor}

\renewcommand{\vr}{\mathbf{r}}
\newcommand{\vf}{\mathbf{f}}

\newcommand{\vk}{\mathbf{k}}

\newcommand{\abs}[1]{\left\vert #1 \right\vert}

\begin{document}

\title{Quantifying translational and bond-orientational order metrics in hyperuniform and nonhyperuniform many-particle systems}
\author{Anirban Mukherjee}
\affiliation{Department of Physics, Princeton University, 
Princeton, New Jersey 08544, USA}

\author{Salvatore Torquato}
\affiliation{Department of Chemistry, Princeton University, 
Princeton, New Jersey 08544, USA}
\affiliation{Department of Physics, Princeton University, 
Princeton, New Jersey 08544, USA}
\affiliation{Princeton Materials Institute, Princeton University, 
Princeton, New Jersey 08544, USA}
\affiliation{Program in Applied and Computational Mathematics, 
Princeton University, Princeton, New Jersey 08544, USA}

\begin{abstract}
Quantifying the degree of order and disorder in many-particle systems remains an
outstanding problem in physics, materials science, and mathematics.
To this end, we consider the translational order metric
$\tau_T$, defined as the squared $L^2$ norm of the total correlation function
$h(\mathbf{r})$, and introduce its bond-orientational analog $\tau_O$,
defined from the weighted total correlation function
$h_{\mathbf{f}}(\mathbf{r})$ using local orientational
weights [S.~Torquato \textit{et al.}, \textit{Phys.\ Rev.\ X} \textbf{16},
011042 (2026)]. Unlike conventional metrics that quantify
either positional order or spatially averaged orientational order, the pair
$(\tau_T,\tau_O)$ places both forms of order on a common two-point statistical
footing, with each sensitive to the amplitude and spatial persistence of
the corresponding correlation function.
The translational order metric $\tau_T$ vanishes for a Poisson point process, whereas the geometry-derived
weights yield a small, construction-dependent positive reference value of
$\tau_O$; each metric diverges in systems possessing
its corresponding form of long-range order. We compute
both metrics for two nonhyperuniform sphere-packing
models in two and three dimensions as functions of packing fraction
$\phi$: (1) equilibrium hard particles along the stable fluid
branches up to near freezing and at selected values of $\phi$ along the
stable crystal branches, and (2) nonequilibrium random sequential addition
(RSA) packings from the dilute regime to just below saturation. For the
nonhyperuniform systems, bond-orientational order remains subdominant along
the equilibrium-fluid and RSA configurations; however, its
magnitude relative to the translational metric increases near the upper end
of the equilibrium-fluid branches, much more strongly in two than in three dimensions, and
the two metrics become comparable along the sampled crystal branches.
At common packing fractions, equilibrium fluids and RSA
packings trace distinct $(\tau_T,\tau_O)$ trajectories,
revealing preparation-dependent differences in structural
order. As a representative hyperuniform family, we study 2D disordered
stealthy hyperuniform (SHU) ground states for
$0<\chi<1/2$, where $\chi$ is the stealthiness parameter.
Within the disordered SHU phase, bond-orientational order
remains subdominant, but its magnitude relative to the
translational metric increases toward the disorder-to-order threshold.
In all three models, $\tau_T$ and $\tau_O$ are positively
correlated beyond the Poisson-reference regime:
$\tau_O$ increases monotonically with $\tau_T$
across the sampled state points. These metrics may
ultimately serve as collective structural coordinates for mapping freezing,
melting, glass formation, jamming, self-assembly, and driven nonequilibrium
transitions, as well as for enhanced sampling, phase classification, and
inverse design.
\end{abstract}

\maketitle

\section{Introduction}

Long-range translational and bond-orientational order are defining
properties of perfect crystals and quasicrystals~\cite{Hansen2013Aug, Torquato2002book,Chaikin1995Jun, Frenkel2023, Steinhardt1987Dec}. In crystals these forms of order arise from periodic
translational symmetry, whereas in quasicrystals they arise from
quasiperiodic translational symmetry that permits noncrystallographic
rotational symmetries~\cite{Levine1984Dec}. Between these perfectly ordered systems and the
perfectly disordered ideal gas lies a broad range of many-particle systems
that lack the full long-range order of perfect crystals and
quasicrystals, but whose degrees of order span a continuous spectrum from
complete disorder to perfect order~\cite{Torquato2000Mar,
Truskett2000Jul}. In principle, their complete statistical
description is contained in the probability density $P_N(\vr^N)$, where
$\vr^N \equiv \{\vr_1,\vr_2,\ldots,\vr_N\}$ specifies the particle
coordinates. However, in practice, it is generally inaccessible for large
$N$, and hence one must settle for reduced but meaningful information, such
as a scalar order metric $\psi_N(\vr^N)$~\cite{Torquato2000Mar,
Kansal2000Sep, Torquato2026May}.

Several scalar translational order metrics have been defined as functionals
of the pair correlation function $g_2(\vr)$ and the total correlation
function $h(\vr)$, related as $h(\vr)=g_2(\vr)-1$. The metric $T$~\cite{Torquato2000Mar,
Truskett2000Jul} measures translational order relative to a perfect face-centered cubic (fcc) lattice at the same number density. The metric
$T^*$~\cite{Truskett2000Jul} integrates $|h(r)|$ over a finite real-space
cutoff. The excess two-particle entropy
$s^{(2)}$~\cite{Nettleton1958Dec, Truskett2000Jul} is the leading pair
contribution to the excess entropy relative to an ideal gas at the same
number density. These metrics have been applied to hard-sphere packings,
Lennard-Jones fluids, and liquid water~\cite{Truskett2000Jul,
Errington2001Jan, Errington2002Oct, Errington2003Feb} to characterize the
degree of translational order in the disordered phase.

On the other hand, the degree of global bond-orientational order is commonly
quantified by one-point functions constructed from locally defined
bond-orientational quantities, such as $\psi_n$ in two dimensions~\cite{Halperin1978Jul,Nelson1979Mar,Keim2007Mar, Mak2006Jun, Thorneywork2017Apr, Han2008Apr} and $Q_l$ in three dimensions~\cite{Steinhardt1983Jul,ReintenWolde1996Jun, Gasser2001Apr, Lechner2008Sep, Kawasaki2010Aug, Leocmach2012Jul}. The global forms $\psi_6$
and $Q_6$ were subsequently used as scalar order metrics for hard-disk and
hard-sphere systems, respectively~\cite{Kansal2000Sep,Rintoul1996Nov}.
Unlike the translational order metrics described above, these global
quantities are functionals of spatial averages of local bond-orientational
weights $\vf_j$ attached to each particle $j$ and therefore do not retain
information about how bond-orientational correlations vary with the
separation between particles.

Here we propose to quantify the degrees of global translational and
bond-orientational order via metrics that are functionals of their respective
two-point correlation functions for statistically homogeneous many-particle
configurations whose particle coordinates are fully specified by their centers
of mass.
The translational order metric $\tau_T$ is the squared $L^2$ norm of the
total correlation function:
\begin{subequations}
\label{eq:tauT}
\begin{align}
\tau_T
&= \frac{1}{D^d}\int_{\mathbb{R}^d} h^2(\vr)\,d\vr,
\label{eq:tauT_real}\\
&= \frac{1}{(2\pi)^dD^d}\int_{\mathbb{R}^d}
\widetilde{h}^{\,2}(\vk)\,d\vk,
\label{eq:tauT_fourier}
\end{align}
\end{subequations}
where $D$ is a characteristic length scale and $\widetilde{h}(\vk)$ is the
Fourier transform of the total correlation function $h(\vr)$.  For square-integrable correlation
functions, the two representations of this metric are equivalent by
Parseval's theorem~\cite{Torquato2015May} and enable accurate
characterization of short-range translational correlations in direct space
[Eq.~\eqref{eq:tauT_real}] and long-wavelength translational correlations in
Fourier space [Eq.~\eqref{eq:tauT_fourier}]~\cite{Lomba2017Dec}~\footnote{In
Ref.~\cite{Lomba2017Dec}, it is shown that the excess
entropy~\cite{Truskett2000Jul} is clearly related to the $\tau_T$ order metric.
The advantage of $\tau_T$ over excess entropy is that the former has both
direct- and Fourier-space representations, which enable accurate capture of
both short- and long-wavelength translational correlations.}. This metric was introduced in Ref.~\cite{Torquato2015May} and
subsequently applied to quantify translational order in several
disordered systems~\cite{Zhang2016Nov,Klatt2019Feb}. The translational metric vanishes for a Poisson point process but
diverges for perfect crystals and quasicrystals in the thermodynamic
limit. For finite periodic or quasiperiodic systems, $\tau_T$ grows with system size
$L$, and
its rate of growth can be used to distinguish the degree of order
among these ordered structures~\cite{Torquato2019Mar,Klatt2020Mar}.

As a bond-orientational analog of $\tau_T$, we introduce the
metric $\tau_O$, defined as
\begin{subequations}
\label{eq:tauO}
\begin{align}
\tau_O
&= \frac{1}{D^d\abs{\vf}_{\rm ref}^4}
\int_{\mathbb{R}^d}\abs{h_{\vf}(\vr)}^2\,d\vr,
\label{eq:tauO_real}\\
&= \frac{1}{(2\pi)^dD^d\abs{\vf}_{\rm ref}^4}
\int_{\mathbb{R}^d}\abs{\widetilde{h}_{\vf}(\vk)}^2\,d\vk,
\label{eq:tauO_fourier}
\end{align}
\end{subequations}
where $h_{\vf}(r)$ is the bond-orientationally weighted total correlation
function~\cite{Torquato2026Mar}, and $\abs{\vf}_{\rm ref}$ is the magnitude
of the local orientational weight in a compatible perfect reference crystal.
The normalization by $\abs{\vf}_{\rm ref}^4$ makes
$\tau_O$ invariant under any uniform rescaling of the bond-orientational
weights. As with $\tau_T$, the equivalent direct-space
[Eq.~\eqref{eq:tauO_real}] and Fourier-space
[Eq.~\eqref{eq:tauO_fourier}] representations of $\tau_O$ capture both
short- and long-wavelength bond-orientational correlations. Unlike
$\tau_T$, the bond-orientational metric can possess a nonzero Poisson
baseline because the construction of the local bond-orientational
weights depends on the identification of geometric neighbors, which
can be shared by nearby particles even in an otherwise uncorrelated
point pattern. For perfect crystals and quasicrystals, Bragg peaks in the
bond-orientational structure factor cause $\tau_O$ to diverge in the
thermodynamic limit.

The order metrics defined by relations~\eqref{eq:tauT}
and~\eqref{eq:tauO} place translational and bond-orientational order
on a common two-point footing (i.e., each metric is constructed from
its respective two-point correlation function), which heretofore has
not been done. We compute the order metrics for models that span from typical nonhyperuniform systems to hyperuniform ones. As representatives of nonhyperuniform models, which do not anomalously suppress infinite-wavelength density fluctuations~\cite{Torquato2003Oct}, we study equilibrium hard-particle configurations~\cite{Hansen2013Aug} and nonequilibrium random sequential addition (RSA) packings~\cite{Zhang2013Nov} in two and three dimensions, using the packing fraction $\phi$ as the control parameter. The equilibrium configurations span the stable fluid branches up to just
short of freezing and are sampled at selected values along the stable crystal
branches, whereas the RSA packings extend from the dilute regime to just
below saturation. Hyperuniform systems anomalously suppress infinite-wavelength density fluctuations, as expressed by the vanishing of the structure factor in the small-wave-number limit, $S(\vk)\to0$ as $|\vk|\to0$~\cite{Torquato2003Oct}
(see Sec.~\ref{subsec:hyperuniformity} for details). Stealthy hyperuniform systems satisfy the stronger condition $S(\vk)=0$ for $0<|\vk|\leq K$~\cite{Batten2008Aug}, where $K$ is the stealthy cutoff wave number~\cite{Torquato2015May}. The dimensionless stealthiness parameter is $\chi = M(K) / [d(N-1)]$, where $M(K)$ is the number of independently constrained wave vectors. As a representative hyperuniform model, we study two-dimensional (2D) SHU ground-state configurations in the disordered regime $0<\chi<1/2$, where $\chi=1/2$ marks the disorder-to-order threshold.

We select these models to determine how global translational and
bond-orientational order develop along qualitatively different routes to
structural organization. Equilibrium hard particles provide crystallizing
nonhyperuniform systems with dimension-dependent~\cite{Hansen2013Aug, Bernard2011Oct} ordering pathways. RSA
packings provide nonequilibrium nonhyperuniform systems that become
kinetically arrested without crystallizing~\cite{Zhang2013Nov}. Disordered SHU ground-state configurations allow us to examine how the two global
order metrics vary within the disordered SHU phase as $\chi$ is
increased~\cite{Torquato2015May}. We restrict the SHU case
to two dimensions as a representative case because the disorder-to-order
threshold at $\chi=1/2$ holds in both two and three dimensions~\cite{Torquato2015May}. Antihyperuniform disordered point
configurations~\cite{Torquato2021May} are not considered here.

In all cases, $\tau_T$ and $\tau_O$ are positively correlated across the
sampled state points, meaning that $\tau_O$ increases monotonically with
$\tau_T$ as the relevant control parameter is varied. Along the
equilibrium fluid branches of the sphere packings and within the disordered SHU
phase, bond-orientational order remains subdominant to translational
order throughout the sampled ranges.
As the 2D hard-disk liquid-hexatic coexistence interval is approached
from below~\cite{Bernard2011Oct}, the sensitivity of $\tau_O$ to $\phi$
increases sharply. A similar, but weaker, increase in the sensitivity of
$\tau_O$ to $\chi$ occurs as the nominal SHU disorder-to-order threshold
is approached.
Along the sampled equilibrium crystal branches of the sphere packings,
the two forms of order become comparable in magnitude. For RSA
packings, which undergo no ordering transition, bond-orientational order
remains subdominant throughout the full range of packing fractions
studied.

In all systems and control-parameter ranges sampled here, $\tau_T$ is
larger than $\tau_O$ and therefore provides a natural reference scale
for comparing their relative magnitudes. Accordingly, we use the ratio
$\tau_O/\tau_T$ to distinguish regimes in which bond-orientational order
is subdominant, $\tau_O/\tau_T\ll 1$, from those in which the two forms
of order are comparable, $\tau_O/\tau_T=\mathcal{O}(1)$. Along the
sampled crystal branches, $\tau_O/\tau_T\approx0.5$--$1$. This ratio
quantifies relative magnitude and is distinct from the positive
correlation between the metrics, which refers to their monotonic
covariation along a control-parameter path. The observed hierarchy
$\tau_O<\tau_T$ is specific to the systems sampled here and is not a
general constraint. For example, in a hexatic phase, where
bond-orientational order is quasi-long-ranged while translational order
remains short-ranged~\cite{Nelson1979Mar,Torquato2026Mar}, $\tau_O$ may
exceed $\tau_T$. More broadly, the pair $(\tau_T,\tau_O)$ may provide
complementary structural objectives for future inverse-design studies
seeking prescribed combinations of translational and bond-orientational
order~\cite{Torquato2022LocalOrder,Shi2025Construction}.

The remainder of this paper is organized as follows.
In Sec.~\ref{sec:background} we provide background
definitions and introduce $\tau_T$ and $\tau_O$.
Section~\ref{sec:models} describes the model systems
and computational methods. In Sec.~\ref{sec:results}
we present and discuss our results. Concluding remarks
are given in Sec.~\ref{sec:conclusions}.

\section{Background and Definitions}
\label{sec:background}

\subsection{Pair statistics and the structure factor}
\label{subsec:pair}

For an $N$-particle system in $d$-dimensional Euclidean space
$\mathbb{R}^d$, with configuration
$\vr^N\equiv(\vr_1,\ldots,\vr_N)$, the reduced $n$-particle density
functions are obtained from the configurational probability density
$P_N(\vr^N)$ by integrating over all but $n$ particle
coordinates~\cite{Torquato2002book,Hansen2013Aug}:
\begin{equation}
\label{eq:rho_n}
\rho_n(\vr^n)
=
\frac{N!}{(N-n)!}
\int \cdots \int
P_N(\vr^N)\,
d\vr_{n+1}\cdots d\vr_N.
\end{equation}
The corresponding $n$-particle correlation functions are
\begin{equation}
g_n(\vr^n)
=
\frac{\rho_n(\vr^n)}{\rho^n},
\end{equation}
where $\rho$ is the number density. For statistically homogeneous systems,
the pair correlation function is
\begin{equation}
g_2(\vr_{12})
=
\frac{\rho_2(\vr_{12})}{\rho^2},
\end{equation}
where $\vr_{12}=\vr_2-\vr_1$. The total correlation function,
\begin{equation}
h(\vr)
=
g_2(\vr)-1,
\label{eq:hr}
\end{equation}
measures the deviation of the pair statistics from those of an uncorrelated
system. For statistically isotropic systems, $g_2$ and $h$ depend only on
$r=|\vr|$, and $g_2(r)$ is the radial distribution function.

We define the Fourier transform of a function $f(\vr)$ as
\cite{Torquato2021May}
\begin{equation}
\tilde{f}(\vk)
=
\int_{\mathbb{R}^d}
f(\vr)\,e^{-i\vk\cdot\vr}\,d\vr,
\end{equation}
with inverse
\begin{equation}
f(\vr)
=
\frac{1}{(2\pi)^d}
\int_{\mathbb{R}^d}
\tilde{f}(\vk)\,e^{i\vk\cdot\vr}\,d\vk.
\end{equation}
The structure factor is related to the Fourier transform of the total
correlation function by~\cite{Hansen2013Aug,Torquato2002book}
\begin{equation}
\label{eq:structure_factor}
S(\vk)
=
1+\rho\,\tilde{h}(\vk).
\end{equation}
It is proportional to the scattering intensity and is directly accessible
in scattering experiments~\cite{Hansen2013Aug}. For statistically isotropic
systems, $S(\vk)$ depends only on $k=|\vk|$. An ideal gas has $h(r)=0$ and
$S(k)=1$, whereas a perfect crystal exhibits Bragg peaks at reciprocal
lattice vectors.

\subsection{Translational order metric}
\label{subsec:tauT}

In statistically isotropic systems, $h(\vr)$ and
$\tilde{h}(\vk)$ become functions only of the magnitudes $r=|\vr|$ and
$k=|\vk|$, respectively, and we can write $\tau_T$ in
Eq.~\eqref{eq:tauT} as
\begin{subequations}
\label{eq:tauT_iso}
\begin{align}
\tau_T
&=
\frac{s_1(1)}{D^d}
\int_0^\infty h^2(r)\,r^{d-1}\,dr,
\label{eq:tauT_iso_real}\\
&=
\frac{s_1(1)}{(2\pi)^dD^d}
\int_0^\infty \widetilde{h}^{\,2}(k)\,k^{d-1}\,dk,
\label{eq:tauT_iso_fourier}
\end{align}
\end{subequations}
where
$s_1(1)
= 2\pi^{d/2} / \Gamma(d/2)$
is the surface area of the unit sphere in $d$ dimensions. For an ideal gas,
or spatially uncorrelated Poisson point process, $h(r)=0$, equivalently
$\tilde{h}(k)=0$, and hence $\tau_T=0$. Thus, deviations of $\tau_T$ from
zero measure the degree of translational order relative to the fully uncorrelated case.

For comparison with the numerical results, we use three analytical references.
For equilibrium and RSA hard-particle systems, the dilute-limit pair correlation
function is $g_2(r)=\Theta(r-2R)+O(\phi)$, where $R$ is the particle
radius~\cite{Torquato2002book}. Substitution of $h(r)=g_2(r)-1$ into
Eq.~\eqref{eq:tauT_iso_real} gives the dilute-limit reference
$\tau_T^{\rm dilute}=2^d\phi$. For three-dimensional (3D) equilibrium hard
spheres, $\tau_T^{\rm PY}$ is obtained analogously using the Percus-Yevick
approximation~\cite{Torquato2002book}. For 2D disordered SHU ground states,
using the choice $D=\rho^{-1/d}$ adopted in this work, the small-$\chi$
prediction is $\tau_T=4\chi+O(\chi^2)$ as
$\chi\to0$~\cite{Torquato2015May}.

For the disordered systems considered here, $h\in L^2$, and the integrals in
Eqs.~\eqref{eq:tauT_iso_real} and~\eqref{eq:tauT_iso_fourier}
are well defined in the thermodynamic limit. However, 
in the crystal phases, quasi-long-range translational order in two dimensions and long-range translational order in three dimensions imply that $h\notin L^2$ in the
thermodynamic limit~\cite{Halperin1978Jul,Nelson1979Mar,
Bernard2011Oct,Klatt2020Mar}. We therefore evaluate the
cutoff-dependent quantity $\tau_T(K_c;L)$,
\begin{equation}
\tau_T(K_c; L)
\equiv
\frac{s_1(1)}{(2\pi)^dD^d}
\int_0^{K_c}
\tilde{h}^{\,2}(k)\,k^{d-1}\,dk,
\label{eq:tauT_Kc}
\end{equation}
where $K_c$ is an imposed radial wave number cutoff and
\begin{equation}
\tilde{h}(k)
=
\frac{1}{\Omega}
\int_{\Omega}
\tilde{h}(\vk)\,d\Omega
\end{equation}
denotes the angular average at fixed $k$, with $\Omega=s_1(1)$.
At fixed system size $L$, $\lim_{K_c\to\infty}\tau_T(K_c;L)=\tau_T(L),$
whereas the cutoff-converged value $\tau_T(L)$ remains system-size
dependent and diverges as $L\to\infty$~\cite{Torquato2019Mar,Klatt2020Mar}.
Here $\tau_T(L)$ denotes the finite-system equilibrium
quantity at fixed $L$, whereas its numerical estimate is obtained from
the shell- and, when available, ensemble-averaged structure factor.

\subsection{Hyperuniformity and the number variance}
\label{subsec:hyperuniformity}

A statistically homogeneous and isotropic many-particle system is said to be
\textit{hyperuniform} if its local number variance $\sigma^2(R)$ within a
spherical observation window of radius $R$ grows more slowly than the window
volume in the large-$R$ limit~\cite{Torquato2003Oct}. Equivalently, the
structure factor satisfies
\begin{equation}
\lim_{k\to0}
S(k) = 0.
\label{eq:hyperuniformity_def}
\end{equation}
Thus, hyperuniform systems suppress infinite-wavelength density
fluctuations. If the structure factor vanishes near the origin as
\begin{equation}
S(k)
\sim
k^\alpha
\qquad
(k\to0,\ \alpha>0),
\end{equation}
then the large-$R$ scaling of the local number variance is
\begin{equation}
\sigma^2(R)
\sim
\begin{cases}
R^{d-1},
& \alpha>1
\quad \text{(class~I)},
\\
R^{d-1}\ln R,
& \alpha=1
\quad \text{(class~II)},
\\
R^{d-\alpha},
& 0<\alpha<1
\quad \text{(class~III)}.
\end{cases}
\end{equation}
Classes I and III represent the strongest and weakest forms of
hyperuniformity, respectively. Perfect crystals and disordered SHU ground
states provide ordered and disordered examples of class I hyperuniformity.
By contrast, typical nonhyperuniform systems have
$\sigma^2(R)\sim R^d$, corresponding to a positive finite limit of $S(k)$
as $k\to0$. Antihyperuniform systems, for which $\sigma^2(R)$ grows faster
than $R^d$, are not considered here.

\subsection{Weighted bond-orientational correlations}
\label{subsec:weightedBOO}

We use the weighted-particle formalism of
Torquato et al.~\cite{Torquato2026Mar} to describe bond-orientational
correlations. Each particle $j$ at position $\vr_j$ carries a weight $\vf_j$
that encodes its local bond-orientational environment. A bond is the
center-to-center connection between a particle and one of its prescribed
nearest neighbors, whose construction is specified in
Sec.~\ref{subsec:computational}.

In two dimensions we use the sixfold bond-orientational order parameter as
a complex scalar weight:
\begin{equation}
f_j
\equiv
\psi_6(j)
=
\frac{1}{N_j}
\sum_{k=1}^{N_j}
\exp(i6\theta_{jk}),
\label{eq:psi6}
\end{equation}
where $N_j$ is the number of prescribed nearest neighbors of particle $j$,
and $\theta_{jk}$ is the angle, measured relative to a fixed reference axis,
of the bond joining particle $j$ to its $k$th neighbor. In three dimensions we use the complex vector weight
\begin{equation}
\vf_j
=
\{Q_{6m}(j)\}_{m=-6}^{6},
\label{eq:f3D}
\end{equation}
where
\begin{equation}
Q_{lm}(j)
\equiv
\sqrt{\frac{4\pi}{2l+1}}\,
\frac{1}{N_j}
\sum_{k=1}^{N_j}
Y_{lm}(\theta_{jk},\varphi_{jk}),
\label{eq:Q6m}
\end{equation}
and $Y_{lm}(\theta,\varphi)$ is the spherical harmonic of degree $l$ and
order $m$. We choose $l=6$ because the corresponding order parameter $Q_6$ is
known to detect fcc bond-orientational ordering~\cite{Torquato2000Mar}. The normalizations in Eqs.~\eqref{eq:psi6}
and~\eqref{eq:Q6m} ensure that the magnitude of the local
orientational weight does not exceed unity in either dimension.

For statistically homogeneous weighted particle systems, let $p(\vf)$
denote the singlet probability density of the weights and
$\rho_2(\vr;\vf_1,\vf_2)$ the two-particle probability density at separation
$\vr$. Following Ref.~\cite{Torquato2026Mar}, the generalized pair
correlation function is
\begin{equation}
g_{2;\vf}(\vr;\vf_1,\vf_2)
\equiv
\frac{\rho_2(\vr;\vf_1,\vf_2)}{\rho^2}.
\label{eq:g2f_general}
\end{equation}
In the absence of long-range order,
$g_{2;\vf}(\vr;\vf_1,\vf_2)\to p(\vf_1)p(\vf_2)$ as
$\abs{\vr}\to\infty$. The weight-averaged pair correlation function is
\begin{equation}
g_{2;\vf}(\vr)
\equiv
\iint
(\vf_1^\dagger\bullet\vf_2)\,
g_{2;\vf}(\vr;\vf_1,\vf_2)\,
d\vf_1\,d\vf_2,
\label{eq:g2f}
\end{equation}
where $\bullet$ denotes an inner product. For the weights used here,
\begin{equation}
\vf_1^\dagger\bullet\vf_2
=
\begin{cases}
\psi_6^*(1)\psi_6(2),
& d=2,
\\[2pt]
\displaystyle
\sum_{m=-6}^{6}Q_{6m}^*(1)Q_{6m}(2),
& d=3.
\end{cases}
\label{eq:weight_inner_product}
\end{equation}
This inner product measures the similarity of the local bond-orientational
environments and is invariant under a global rotation. The corresponding
weighted total correlation function is
\begin{equation}
h_{\vf}(\vr)
\equiv
g_{2;\vf}(\vr)-\abs{\langle\vf\rangle}^2,
\label{eq:hf}
\end{equation}
where $\langle\vf\rangle=\int\vf\,p(\vf)\,d\vf$.
From Eq.~\eqref{eq:g2f}, particle-exchange symmetry implies that
$h_{\vf}(\vr)$ satisfies
\begin{equation}
h_{\vf}(-\vr)
=
h_{\vf}^*(\vr).
\label{eq:hf_hermitian}
\end{equation}
For statistically isotropic systems,
$h_{\vf}(-\vr)=h_{\vf}(\vr)=h_{\vf}(r)$, and
Eq.~\eqref{eq:hf_hermitian} therefore implies that $h_{\vf}(r)$ is real.

Although $h_{\vf}(\vr)$ is a two-point correlation function of the weighted
field, the geometry-derived weights depend on local particle neighborhoods
and therefore encode higher-order positional information from the underlying
unweighted configuration.

\subsection{Bond-orientational order metric}
\label{subsec:tauO}

The bond-orientational order metric $\tau_O$ is defined in
Eqs.~\eqref{eq:tauO_real} and \eqref{eq:tauO_fourier}. The normalization by
$\abs{\vf}_{\rm ref}^{4}$ makes $\tau_O$ invariant under a uniform rescaling
of the orientational weights, where $\abs{\vf}_{\rm ref}$ is the magnitude
of the local bond-orientational weight in the corresponding perfect reference
crystal. In statistically isotropic systems,
$h_{\vf}(\vr)$ and $\tilde{h}_{\vf}(\vk)$ become functions only of the
magnitudes $r=|\vr|$ and $k=|\vk|$, respectively, and are real. We can
therefore write $\tau_O$ in Eq.~\eqref{eq:tauO} as
\begin{subequations}
\label{eq:tauO_iso}
\begin{align}
\tau_O
&=
\frac{s_1(1)}
{D^d\abs{\vf}_{\rm ref}^{4}}
\int_0^\infty
\left[h_{\vf}(r)\right]^2 r^{d-1}\,dr,
\label{eq:tauO_iso_real}\\
&=
\frac{s_1(1)}
{(2\pi)^dD^d\abs{\vf}_{\rm ref}^{4}}
\int_0^\infty
\left[\tilde{h}_{\vf}(k)\right]^2 k^{d-1}\,dk.
\label{eq:tauO_iso_fourier}
\end{align}
\end{subequations}
The square-integrability and convergence conditions for the disordered
systems are the same as for $\tau_T$ in Sec.~\ref{subsec:tauT}, with
$h$ replaced by $h_{\vf}$. 
For the crystal phases considered here,
$h_{\vf}$ inherits the non-square-integrable crystalline correlations
because the bond-orientational weights remain coherently ordered over
long distances in both two and three dimensions~\cite{Nelson1979Mar,Steinhardt1983Jul}. We therefore
report the radial cutoff-dependent quantity
\begin{equation}
\tau_O(K_c;L)
\equiv
\frac{s_1(1)}
{(2\pi)^dD^d\abs{\vf}_{\rm ref}^{4}}
\int_0^{K_c}
\abs{\tilde{h}_{\vf}(k)}^2 k^{d-1}\,dk,
\label{eq:tauO_Kc}
\end{equation}
where
\begin{equation}
\tilde{h}_{\vf}(k)
=
\frac{1}{\Omega}
\int_{\Omega}
\tilde{h}_{\vf}(\vk)\,d\Omega
\end{equation}
is the angularly averaged radial function, with $\Omega=s_1(1)$.
At fixed system size $L$,
$\lim_{K_c\to\infty}\tau_O(K_c;L)=\tau_O(L)$, whereas the
cutoff-converged value $\tau_O(L)$ remains system-size dependent and
diverges as $L\to\infty$. Here $\tau_O(L)$ denotes the finite-system equilibrium
quantity at fixed $L$, whereas its numerical estimate is obtained from
the shell- and, when available, ensemble-averaged bond-orientational
structure factor.
For the corresponding perfect reference crystal,
$\vf_j=\vf_{\rm ref}$ at every particle site. It then follows from
Eqs.~\eqref{eq:g2f} and~\eqref{eq:hf} that the angularly averaged functions
$h_{\vf}(r)$ and $h(r)$ are related by
$h_{\vf}(r)=\abs{\vf}_{\rm ref}^{2}h(r)$ and, consequently,
$\tilde{h}_{\vf}(k)=\abs{\vf}_{\rm ref}^{2}\tilde{h}(k)$. Therefore,
comparing Eqs.~\eqref{eq:tauT_Kc} and~\eqref{eq:tauO_Kc}, we obtain
\begin{equation}
\tau_O(K_c;L)
=
\tau_T(K_c;L)
\label{eq:tau_equal_perfect_crystal}
\end{equation}
for any common radial wave-number cutoff $K_c$ and system size $L$.

\section{Model Systems and Computational Methods}
\label{sec:models}

\subsection{Equilibrium hard-particle systems}
\label{subsec:hard_particles}

We study 2D hard disks and 3D hard spheres
along their stable fluid branches up to just short of freezing and at
selected packing fractions along their stable crystal branches. The control
parameter is the packing fraction
\begin{equation}
    \phi=\rho v_1(R),
\end{equation}
where $\rho$ is the number density, $R$ is the particle radius, and
$v_1(R)=\pi^{d/2}R^d/\Gamma(1+d/2)$ is the volume of a
$d$-dimensional sphere of radius $R$. The target-state configurations are
equilibrated in the canonical ensemble under periodic boundary conditions
using the preparation protocols described below. Before evaluating
$\tau_T$ and $\tau_O$, the particle coordinates, particle radii, and
simulation box are uniformly rescaled to unit number density, $\rho=1$.

We consider both $d=2$ and $d=3$ because their crystallization pathways
differ qualitatively. With increasing packing fraction, 2D
hard disks undergo a first-order liquid-hexatic transition with a narrow
coexistence interval, $0.700\lesssim\phi\lesssim0.716$, followed by a
continuous hexatic-solid transition near
$\phi\approx0.720$~\cite{Bernard2011Oct}. Within the hexatic phase,
bond-orientational correlations are quasi-long-ranged, whereas
translational correlations remain short-ranged. By contrast,
3D hard spheres undergo a first-order fluid-crystal
transition with coexistence between the fluid freezing value
$\phi_F\approx0.494$ and the crystal melting value
$\phi_M\approx0.545$, without an analogous intermediate
phase~\cite{Torquato2002book}. This dimensional contrast allows us to
compare how translational and bond-orientational order develop along the
fluid and crystal branches, away from the transition regions, in systems
with qualitatively different crystallization pathways.

\subsubsection{Fluid branches of equilibrium hard disks and hard spheres}
\label{subsubsec:hard_fluid}

For 2D hard disks, we use systems with $N=10{,}000$
particles over $0.01\leq\phi\leq0.68$. For 3D hard
spheres, we use systems with $N=100{,}000$ particles over
$0.01\leq\phi\leq0.49$. These ranges extend from the dilute regime to
just below the respective freezing transitions.

Direct RSA configurations at the target packing fraction were used as
initial conditions whenever they could be generated efficiently. For
$0.01\leq\phi<0.50$ in two dimensions and
$0.01\leq\phi\leq0.19$ in three, RSA configurations were
generated directly at the target packing fraction and subsequently
equilibrated. For $0.50\leq\phi\leq0.68$ in two dimensions, an RSA
configuration at $\phi=0.40$ was used as the initial state, whereas for
$0.20\leq\phi\leq0.49$ in three, an RSA configuration at
$\phi=0.10$ was used. In both cases, the particles were grown
isotropically to the target packing fraction~\cite{skoge2006} before
equilibration. At each state point, $10{,}000$ and $1{,}000$ equilibrated
configurations were sampled for the 2D and 3D systems,
respectively.

\begin{figure}[h]
    \centering
    \includegraphics[width=\linewidth]
    {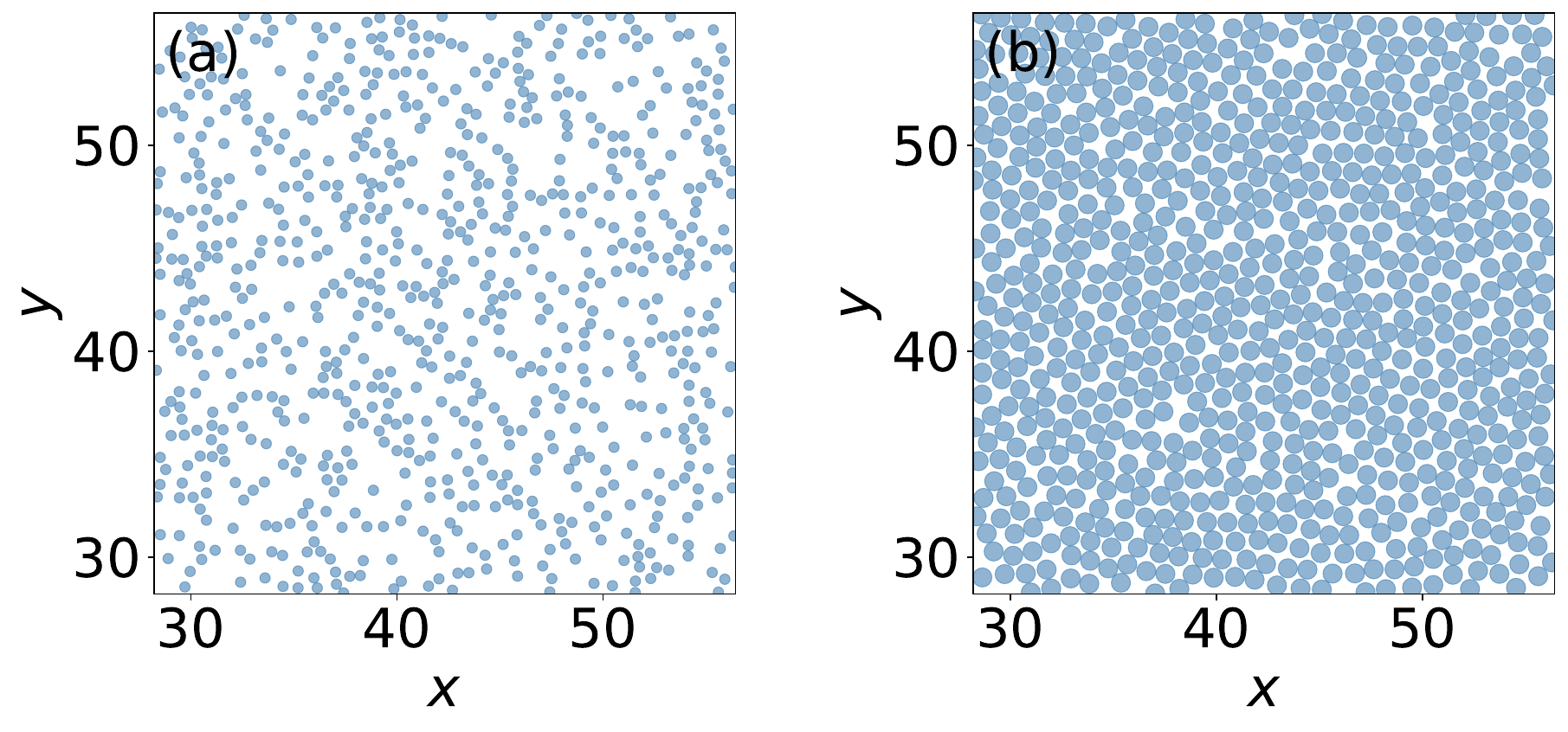}
    \caption{Representative fluid-branch configurations of
    2D equilibrium hard disks at
    (a)~$\phi=0.20$ and (b)~$\phi=0.68$, each displaying the same area
    and containing approximately $800$ disk centers. The
    higher-packing-fraction configuration exhibits stronger short-range
    positional correlations.}
    \label{fig:hard_disk_fluid_configs}
\end{figure}

Representative 2D fluid configurations at low and
near-freezing packing fractions are shown in
Fig.~\ref{fig:hard_disk_fluid_configs}.

\subsubsection{Crystal branches of equilibrium hard disks and hard spheres}
\label{subsubsec:hard_crystal}

\begin{figure}[h]
    \centering
    \includegraphics[width=\linewidth]
    {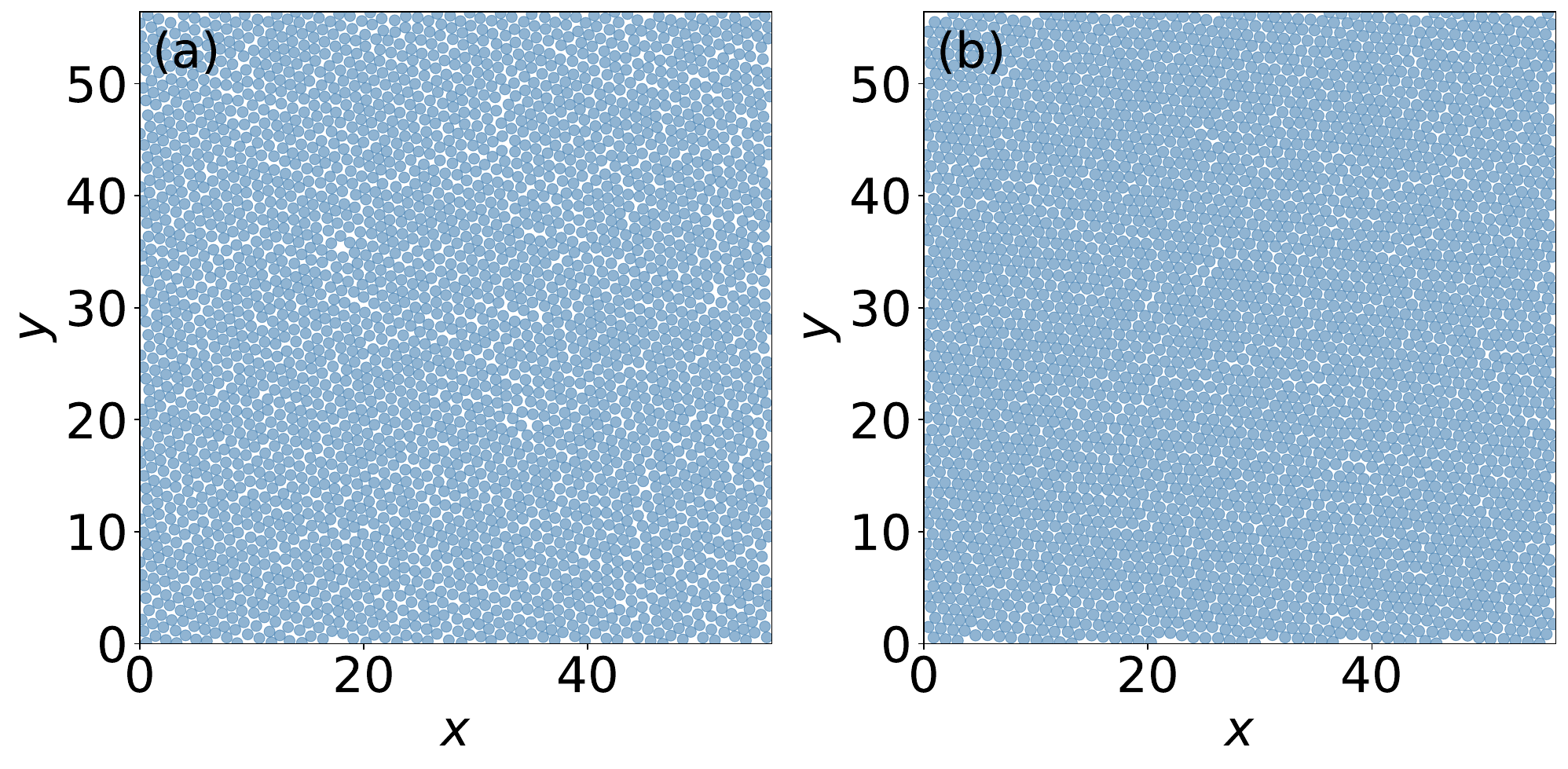}
    \caption{Representative crystal-branch configurations of
    2D equilibrium hard disks at
    (a)~$\phi=0.75$ and (b)~$\phi=0.81$, each displaying the same area
    and containing approximately $3200$ disk centers. The density of
    lattice defects is lower at the higher packing fraction.}
    \label{fig:crystal_configs}
\end{figure}

For 2D hard disks, we use systems with
$N=160{,}000$ particles in a rhombic periodic box compatible with a
triangular crystal. We study the packing fractions
$\phi=0.75$, $0.77$, $0.79$, and $0.81$ along the stable solid branch.
At each target packing fraction, an RSA configuration at $\phi=0.30$
was used as the initial state. The system was isotropically compressed to
the target packing fraction and subsequently equilibrated using a Monte
Carlo algorithm~\cite{Metropolis1953Jun}. Only the final equilibrated
configuration was used for the structural analysis, after verifying that
it contained no polycrystalline grains and exhibited true long-range
bond-orientational correlations. Representative 2D crystal configurations are shown in
Fig.~\ref{fig:crystal_configs}.

For 3D hard spheres, we use systems with
$N=108{,}000$ particles in a cubic periodic box compatible with an fcc
crystal. We study the packing fractions
$\phi=0.56$, $0.60$, $0.65$, and $0.70$ along the stable crystal
branch. At each packing fraction, ten independent trajectories were
initialized from a perfect fcc configuration constructed directly at the
target packing fraction and equilibrated using a Monte Carlo
algorithm~\cite{Metropolis1953Jun}. The final equilibrated configuration from
each trajectory was used for the structural analysis.

\subsection{Random sequential addition packings}
\label{subsec:rsa}

We next consider RSA packings, which provide
nonequilibrium systems that approach saturation without crystallizing.
Starting from an empty periodic simulation box, particles are inserted
randomly and sequentially, with any placement that overlaps a previously
accepted particle rejected. The process continues until the target number
of particles is accepted. At saturation, no additional particle can be inserted without overlap.
Saturated RSA packings are statistically isotropic and nonhyperuniform, with
$\lim_{k\to0}S(k)>0$ in finite spatial
dimensions~\cite{Zhang2013Nov}.

\begin{figure}[!ht]
    \centering
    \includegraphics[width=\linewidth]
    {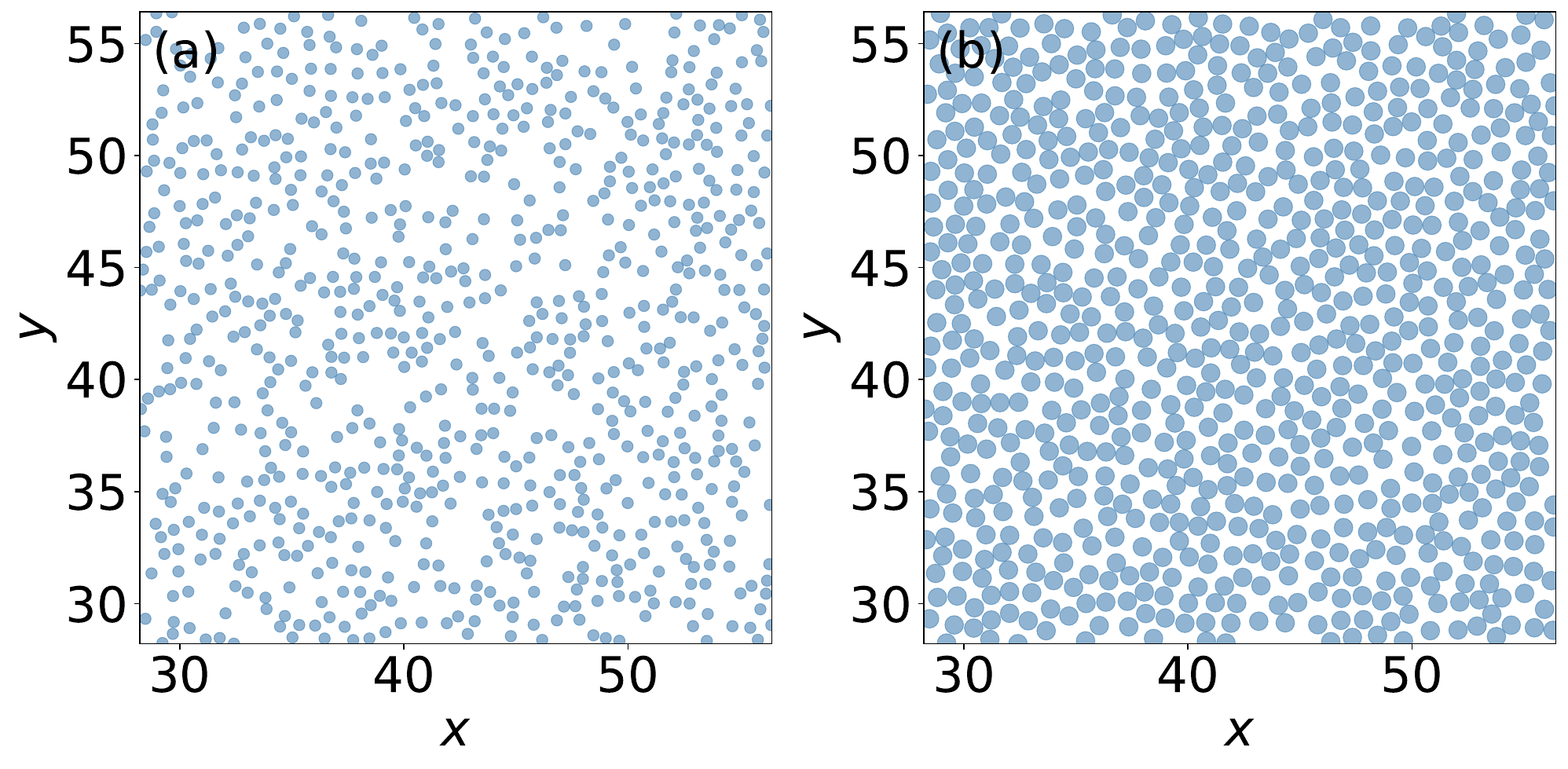}
    \caption{Representative 2D RSA hard-disk configurations at
    (a)~$\phi=0.20$ and (b)~$\phi=0.54$, each displaying the same area
    and containing approximately $800$ disk centers. The
    higher-packing-fraction configuration exhibits stronger short-range
    exclusion-induced correlations.}
    \label{fig:rsa_configs}
\end{figure}

For 2D RSA packings, we use systems with $N=10{,}000$
particles and study packing fractions in the range
$0.01\leq\phi\leq0.54$, extending from the dilute regime to just below
the saturation value $\phi_s\approx0.547$. For 3D RSA
packings, we use systems with $N=100{,}000$ particles and study
$0.01\leq\phi\leq0.37$, extending to just below the saturation value
$\phi_s\approx0.384$~\cite{Zhang2013Nov}.
At each packing fraction, the particle radius was chosen according to the
target value of $\phi$ at unit number density, $\rho=1$, and the particles
were deposited sequentially in a periodic square box in two dimensions or a periodic cubic box in three. We generated $10{,}000$
independent configurations at each state point in two dimensions and
$100$ independent configurations at each state point in three.
Representative 2D RSA configurations at intermediate and
near-saturated packing fractions are shown in
Fig.~\ref{fig:rsa_configs}.

\subsection{Stealthy hyperuniform ground-state configurations}
\label{subsec:shu}

We consider 2D stealthy hyperuniform (SHU) ground-state
configurations generated using the collective-coordinate ground-state
optimization procedure described in
Refs.~\cite{Uche2006Sep,Batten2008Aug,Torquato2015May}. These
configurations satisfy
\begin{equation}
    S(k)=0, \qquad 0<k\leq K,
    \label{eq:SHU_sk}
\end{equation}
so that density fluctuations are completely suppressed over the prescribed
range of wave numbers. The control parameter is the stealthiness parameter
$\chi=M(K)/[d(N-1)]$~\cite{Batten2008Aug}, where $M(K)$ is the number
of independently constrained wave vectors. A larger $\chi$ corresponds to a
larger value of $K$ and hence greater stealthiness. Correspondingly, in real
space, the dimensionless surface-area coefficient governing the asymptotic
number variance decreases as $\chi$ increases, indicating stronger
suppression of large-scale density
fluctuations~\cite{Torquato2015May}.

\begin{figure}[!h]
    \centering
    \includegraphics[width=\linewidth]
    {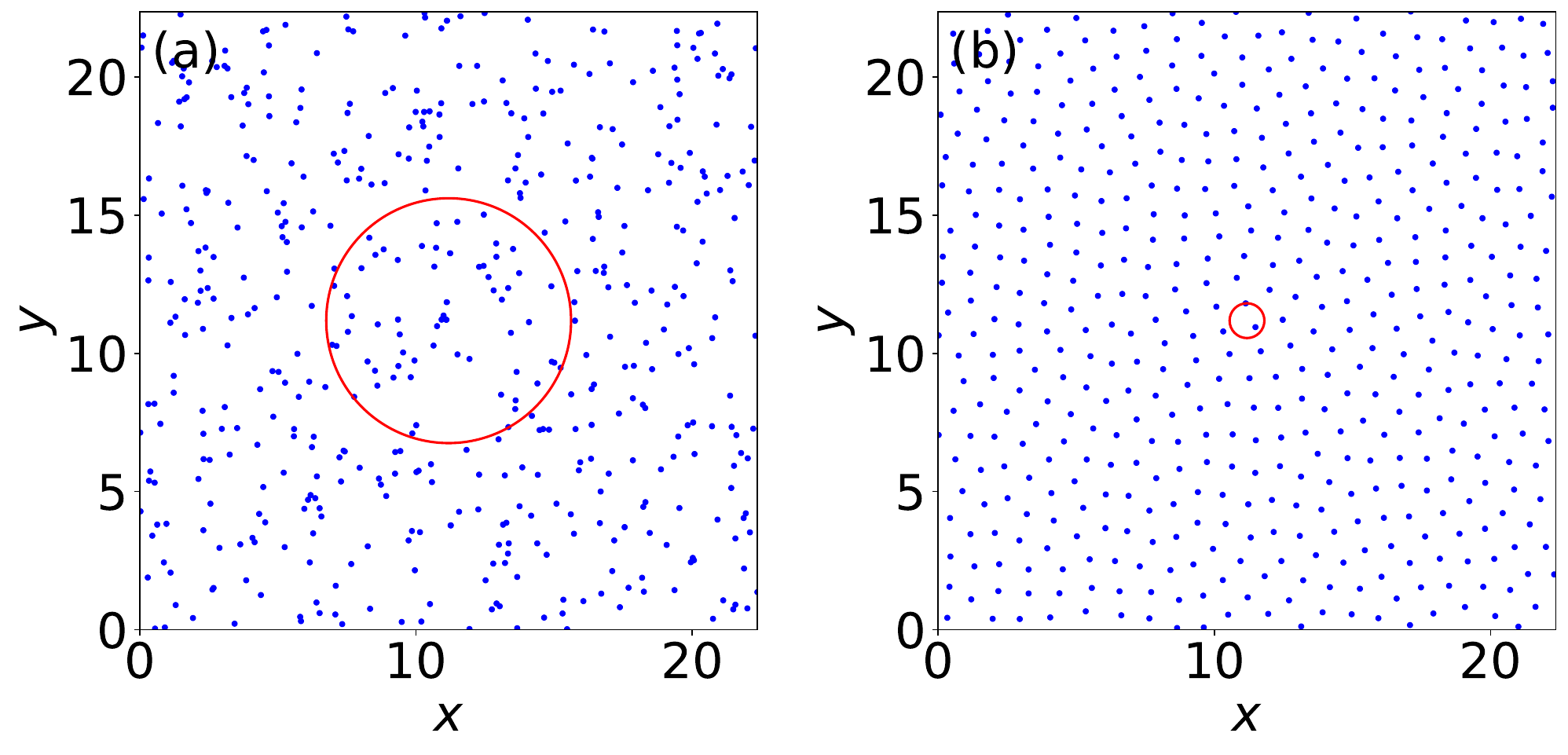}
    \caption{Representative 2D disordered SHU ground-state configurations
    at (a)~$\chi=0.01$ and (b)~$\chi=0.495$. Each panel displays the same
    area and contains approximately $500$ points. The circle indicates the
    stealthy wavelength $\lambda=2\pi/K$. The greater structural regularity
    at the larger value of $\chi$ is visible.}
    \label{fig:SHU_configs_chi_0.01_0.495}
\end{figure}

In the thermodynamic limit, the entropically favored SHU ground states
undergo a disorder-to-order transition at $\chi=1/2$ in both $d=2$ and
$d=3$~\cite{Torquato2015May}. We study only the isotropic
disordered regime in two dimensions as a representative case, using $N=5000$ particles
at unit number density over $0.001\leq\chi\leq0.495$. For each $\chi$, the
stealthy cutoff is
\begin{equation}
    K=4\sqrt{\pi\chi},
\end{equation}
and $4000$ independent configurations are generated from random initial
conditions. The collective-coordinate objective is minimized using L-BFGS
with an energy tolerance of $10^{-16}$.
Representative configurations at $\chi=0.01$ and $\chi=0.495$ are shown in
Fig.~\ref{fig:SHU_configs_chi_0.01_0.495}.

\subsection{Computation of bond-orientational weights and order metrics}
\label{subsec:computational}

The local bond-orientational weights are computed from
Eqs.~\eqref{eq:psi6}--\eqref{eq:Q6m}. Nearest neighbors are defined using
the periodic Delaunay tessellation for all disordered systems and for the
2D crystal branch. For the 3D fcc crystal
branch, we instead use exactly 12 periodic nearest neighbors,
corresponding to the physical first coordination shell, because degeneracies
in the fcc Delaunay tessellation introduce additional neighbors. The
triangular lattice in two dimensions and the fcc lattice in three are the corresponding
thermodynamically stable crystalline structures. In
Eqs.~\eqref{eq:tauO_real} and \eqref{eq:tauO_fourier},
$\abs{\vf}_{\rm ref}$ denotes the magnitude of the local orientational
weight in the compatible perfect reference crystal. For the triangular
crystal, $\psi_6=1$ and hence $\abs{\vf}_{\rm ref}=1$, whereas for the fcc
crystal, using the same 12-nearest-neighbor convention,
\begin{equation}
    \abs{\vf}_{\rm ref}
    =
    \left[
        \sum_{m=-6}^{6}
        \left|Q_{6m}^{\rm fcc}\right|^2
    \right]^{1/2}
    \approx 0.575.
\end{equation}

Throughout this work, all configurations are expressed at unit number
density, $\rho=1$, before evaluating $\tau_T$ and $\tau_O$. The
characteristic length, defined as the mean interparticle spacing, is
therefore
\begin{equation}
    D=\rho^{-1/d}=1.
\end{equation}
The equilibrium fluid branches, RSA packings, and disordered SHU ground
states are statistically isotropic, whereas the crystal branches are
isotropized by angular averaging over reciprocal-space shells.
For the equilibrium fluid branches and RSA packings, we compute $h(r)$
and $h_{\vf}(r)$ directly and evaluate the isotropic direct-space forms in
Eqs.~\eqref{eq:tauT_iso_real} and \eqref{eq:tauO_iso_real}. At every state
point, the cumulative integrals reach plateaus well below the maximum
independent separation allowed by the periodic boundaries, and we report
these plateau values.

For the disordered SHU ground-state configurations and crystal branches, we
evaluate the metrics in reciprocal space. For a periodic simulation cell,
the allowed wave vectors are
\begin{equation}
    \vk=\sum_{i=1}^{d}n_i\mathbf b_i,
\end{equation}
where $\mathbf b_i$ are the reciprocal primitive vectors and $n_i$ are
integers. The reciprocal-space integral is replaced by
\begin{equation}
    \frac{1}{(2\pi)^d}\int d\vk
    \;\longrightarrow\;
    \frac{1}{V}\sum_{\vk},
\end{equation}
where $V$ is the volume of the fundamental cell. Using
Eq.~\eqref{eq:structure_factor}, the isotropized finite-cell translational
metric is
\begin{equation}
    \tau_T(K_c)
    =
    \frac{1}{\rho^2VD^d}
    \sum_{0<k_s\leq K_c}
    m(k_s)
    \left[\overline S(k_s)-1\right]^2 .
    \label{eq:tauT_discrete}
\end{equation}
Here $k_s$ is the wave-number magnitude associated with shell $s$,
$m(k_s)$ is the number of reciprocal vectors in that shell,
$\overline S(k_s)$ is the shell- and ensemble-averaged structure factor,
and $K_c$ is the radial wave-number cutoff.
Equation~\eqref{eq:tauT_discrete} is the finite-cell discrete approximation
to Eq.~\eqref{eq:tauT_Kc}. For the disordered SHU configurations, provided
$L$ is sufficiently large that the omitted finite-size contribution is
negligible and $K_c$ is sufficiently large,
Eq.~\eqref{eq:tauT_discrete} converges to the thermodynamic-limit form in
Eq.~\eqref{eq:tauT_iso_fourier}.

The weighted collective coordinate and bond-orientational structure factor
are
\begin{equation}
    \begin{aligned}
        \mathbf F_{\vf}(\vk)
        &=
        \sum_{j=1}^{N}
        \vf_j\exp(-i\vk\cdot\vr_j),
        \\
        S_O(\vk)
        &=
        \frac{1}{N}
        \left\langle
            \mathbf F_{\vf}^{\dagger}(\vk)
            \bullet
            \mathbf F_{\vf}(\vk)
        \right\rangle .
    \end{aligned}
    \label{eq:SO_def}
\end{equation}
For $\vk\neq\mathbf 0$,
\begin{equation}
    S_O(\vk)-\left\langle\abs{\vf}^2\right\rangle
    =
    \rho\widetilde h_{\vf}(\vk),
\end{equation}
and hence the isotropized finite-cell bond-orientational metric is
\begin{equation}
    \tau_O(K_c)
    =
    \frac{1}
    {\rho^2VD^d\abs{\vf}_{\rm ref}^{4}}
    \sum_{0<k_s\leq K_c}
    m(k_s)
    \left|
        \overline S_O(k_s)
        -\left\langle\abs{\vf}^2\right\rangle
    \right|^2 .
    \label{eq:tauO_discrete}
\end{equation}
Here $\overline S_O(k_s)$ is the shell- and ensemble-averaged
bond-orientational structure factor. The subtraction of
$\langle\abs{\vf}^2\rangle$ removes the self-contribution, in direct
analogy with the subtraction of unity from $S(\vk)$.
Equation~\eqref{eq:tauO_discrete} is the finite-cell discrete approximation
to Eq.~\eqref{eq:tauO_Kc} and, for the disordered SHU ground-state
configurations under the same conditions stated above, converges to the
thermodynamic-limit form in Eq.~\eqref{eq:tauO_iso_fourier}, with $\tau_O$
becoming insensitive to further increases in $K_c$.
For the crystal branches, the dependence of
$\tau_T(K_c)$ and $\tau_O(K_c)$ on the fixed system size $L$ is
implicit. At fixed $L$,
$\lim_{K_c\to\infty}\tau_T(K_c)=\tau_T(L)$ and
$\lim_{K_c\to\infty}\tau_O(K_c)=\tau_O(L)$; these cutoff-converged
values remain system-size dependent and diverge as $L\to\infty$.

For ensemble-averaged data, the statistical uncertainties in $\tau_T$
and $\tau_O$ are estimated separately using leave-one-configuration-out
jackknife standard errors~\cite{Efron1993}. In all cases, their
noise-to-signal ratios are much smaller than unity, and the growth
trends are clear.

\section{Results}
\label{sec:results}

\subsection{Interpretation of the order metrics}
\label{subsec:sensitivity}

For each model we examine how the translational and bond-orientational
order metrics, $\tau_T$ and $\tau_O$, vary along a one-parameter
structural pathway: packing fraction $\phi$ for the equilibrium and RSA
sphere packings, and stealthiness parameter $\chi$ for the SHU ground
states. 
Throughout the results below, the local orientational weight entering
$\tau_O$ is $f_j=\psi_6(j)$ in two dimensions [Eq.~\eqref{eq:psi6}] and
$\vf_j=\{Q_{6m}(j)\}_{m=-6}^{6}$ in three [Eq.~\eqref{eq:f3D}], with the components $Q_{6m}(j)$ defined in
Eq.~\eqref{eq:Q6m}.
We characterize each pathway using three complementary
quantities: the parametric trajectory in the $(\tau_T,\tau_O)$ plane,
the relative magnitude $\tau_O/\tau_T$, and, where useful, the
sensitivities $E_T$ and $E_O$ [see
Eqs.~\eqref{eq:log_response} and~\eqref{eq:log_response_discrete}].
Here a ``positive correlation'' means that $\tau_O$ increases
monotonically with $\tau_T$ along the sampled pathway; it does not
denote a Pearson, Spearman, or other statistical correlation
coefficient. We call bond-orientational order subdominant when
$\tau_O/\tau_T\ll1$ and the two forms of order comparable when this
ratio is of order unity. Because $\tau_O$ has a small positive Poisson
reference while $\tau_T$ vanishes in the Poisson limit, we interpret
$\tau_O/\tau_T$ only beyond the Poisson-reference regime
(see Appendix~\ref{app:poisson_reference}).

We define the sensitivity of each of the two order metrics with respect to
the control parameter as
\begin{equation}
E_\alpha
\equiv
\frac{d\ln\tau_\alpha}{d\ln\lambda} =
\frac{\lambda}{\tau_\alpha}
\frac{d\tau_\alpha}{d\lambda},
\qquad
\alpha=T,O,
\label{eq:log_response}
\end{equation}
where $\lambda=\phi$ for the equilibrium and RSA sphere packings and
$\lambda=\chi$ for the SHU ground states. Thus, a $1\%$ change in $\lambda$
produces approximately an $E_\alpha\%$ change in $\tau_\alpha$.
For discretely sampled data,
we use the centered finite-difference estimate
\begin{equation}
E_{\alpha,i}
\approx
\frac{
\left(\tau_{\alpha,i+1}-\tau_{\alpha,i-1}\right)
/\tau_{\alpha,i}
}{
\left(\lambda_{i+1}-\lambda_{i-1}\right)/\lambda_i
}.
\label{eq:log_response_discrete}
\end{equation}

Representative finite-size diagnostics for the sampled
fluid and disordered phases are presented in
Appendix~\ref{app:representative_correlations} using the
control-parameter values closest to their respective ordering thresholds or saturation limits. The 2D and 3D hard-particle correlation functions decay
on length scales much smaller than the size of the simulation boxes. For the 2D and 3D equilibrium crystal branches,
the small-$k$ behavior of $S(k)$ is consistent with the
free-volume-theory prediction~\cite{Wang2024Aug,Torquato2026May}. In the SHU ensemble,
sufficiently many allowed wave vectors lie within the stealthy region
$0<k\leq K$, where $S(k)=0$, while $S_O(k)$ remains finite down to the
smallest nonzero wave number, $k_{\min}=2\pi/L$, where $L$ is the
simulation box side length, and exhibits no apparent long-wavelength
divergence. These observations indicate that the system sizes used
are sufficient for the reported fluid and disordered-state metrics.

\subsection{Equilibrium sphere packings}
\label{subsec:results_equilibrium}

\begin{figure}[!h]
\centering
\includegraphics[width=\linewidth]
{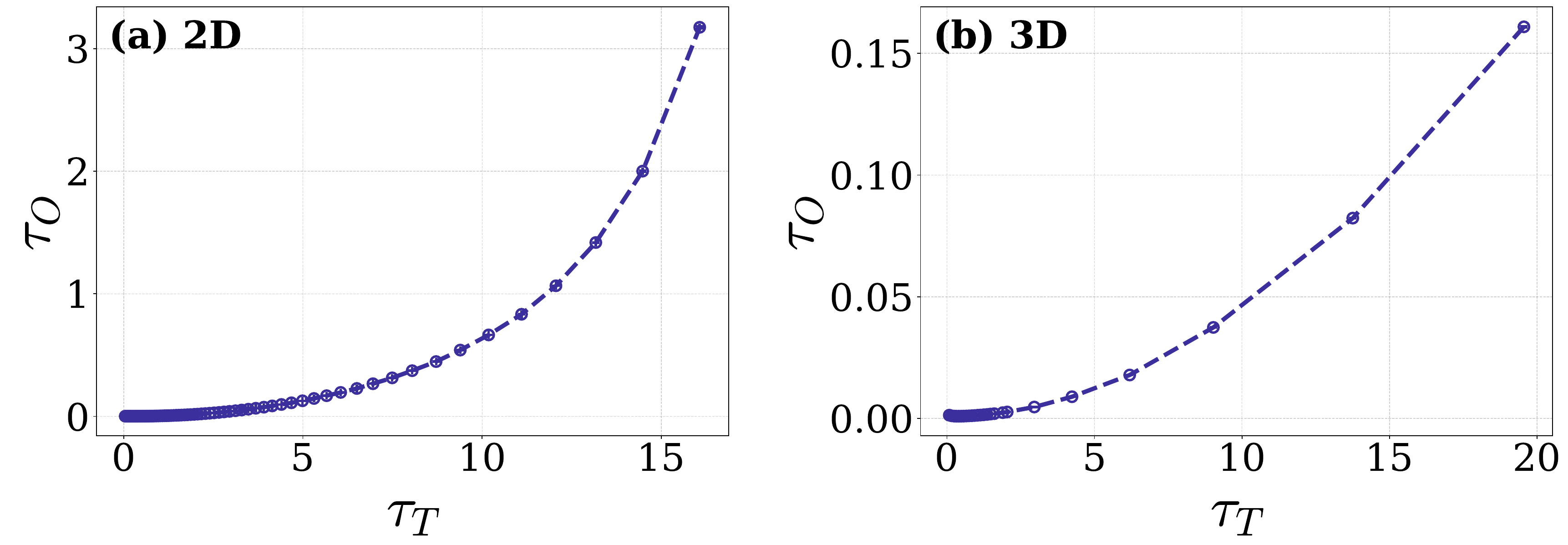}
\caption{Parametric relation between $\tau_O$ and $\tau_T$ along
the stable fluid branches of (a) 2D equilibrium hard disks and
(b) 3D equilibrium hard spheres. In both dimensions, $\tau_O$
increases monotonically with $\tau_T$ along the sampled branches.}
\label{fig:eq_fluid_parametric}
\end{figure}

Across all sampled equilibrium fluid and crystal branches, $\tau_O$
increases monotonically with $\tau_T$; the fluid-branch trajectories
are shown in Fig.~\ref{fig:eq_fluid_parametric}. The magnitude and rate
of this increase, however, depend strongly on dimensionality and phase.

\subsubsection{Stable fluid branch of 2D hard disks}
\label{subsubsec:results_eq_fluid_2D}

\begin{figure}[!h]
\centering
\includegraphics[width=\linewidth]{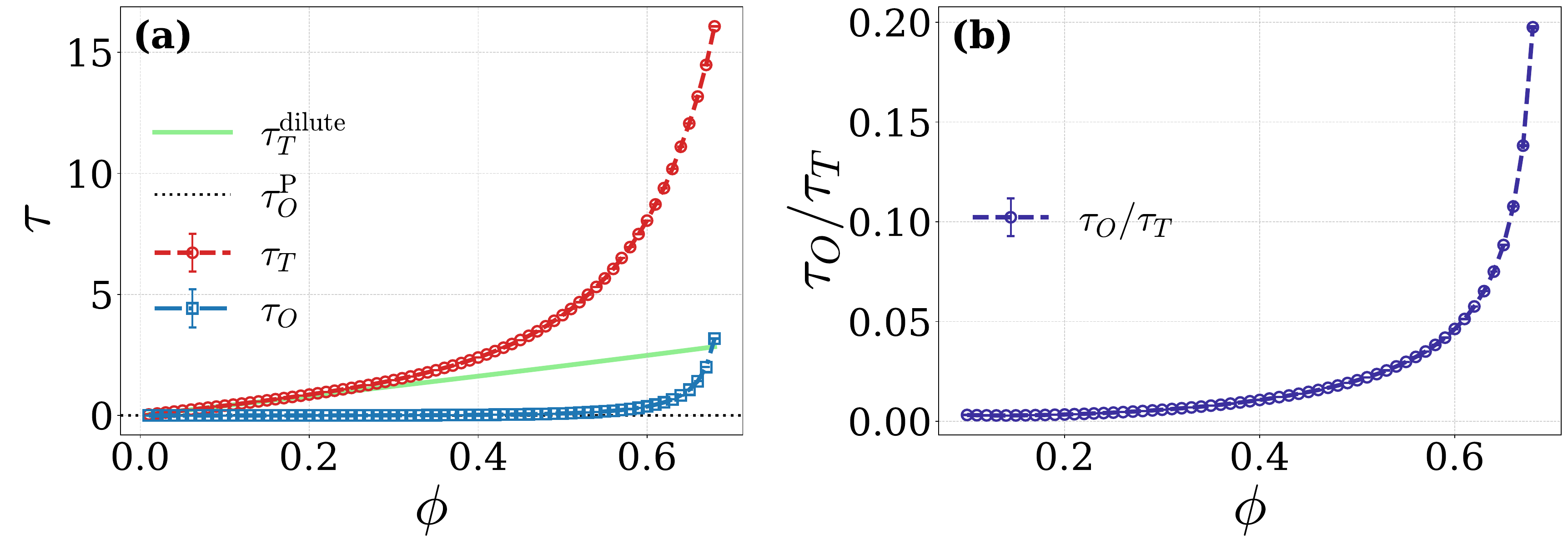}
\caption{Along the stable fluid branch of 2D equilibrium hard
disks, (a) $\tau_T$ and $\tau_O$ as functions of $\phi$, together
with the dilute-limit prediction $\tau_T^{\rm dilute}=4\phi$ for
$\tau_T$, and (b) the ratio $\tau_O/\tau_T$. Although $\tau_O$
remains subdominant, its magnitude relative to $\tau_T$ increases
substantially toward the upper end of the branch.}
\label{fig:eq_hd_fluid}
\end{figure}

\begin{figure}[!h]
\centering
\includegraphics[width=\linewidth]
{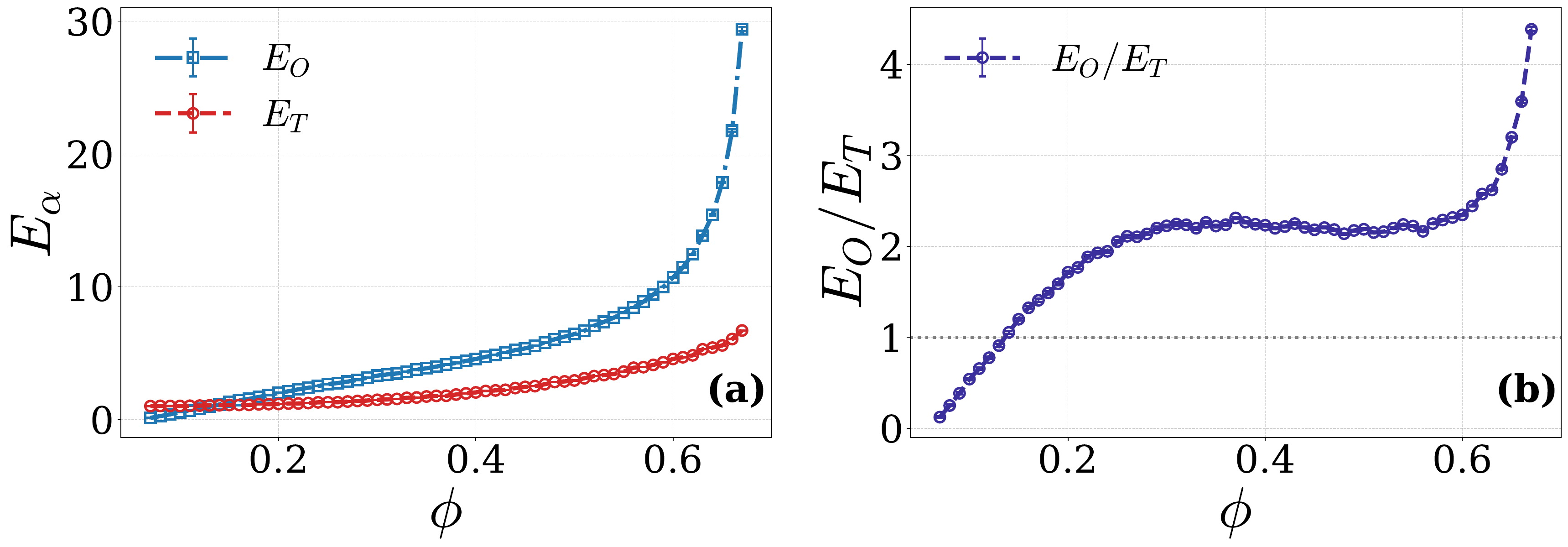}
\caption{Sensitivities $E_T$ and $E_O$ and their ratio
$E_O/E_T$ along the stable fluid branch of 2D equilibrium hard
disks. The larger fractional growth of $\tau_O$, which becomes more
pronounced near the upper end of the fluid branch, explains the
increase in $\tau_O/\tau_T$.}
\label{fig:eq_hd_fluid_fractional_response}
\end{figure}

Along the 2D fluid branch, $\tau_T$ agrees with its dilute-limit
prediction at low $\phi$ and departs from it as hard-core correlations
become appreciable [Fig.~\ref{fig:eq_hd_fluid}(a)].
Bond-orientational order remains subdominant throughout the sampled
branch, but its magnitude relative to $\tau_T$ grows substantially approaching the freezing point:
$\tau_O/\tau_T$ increases from approximately $3.4\times10^{-3}$ at
$\phi=0.20$ to $0.198$ at $\phi=0.68$
[Fig.~\ref{fig:eq_hd_fluid}(b)].

This increase results from the sharply increasing sensitivity of
$\tau_O$ close to the freezing point. Specifically, $E_O/E_T$ increases
overall from approximately $1.72$ at $\phi=0.20$ to $4.38$ at
$\phi=0.67$ [Fig.~\ref{fig:eq_hd_fluid_fractional_response}]. Thus,
near the upper end of the sampled stable-fluid branch, orientational
correlations strengthen fractionally more rapidly than translational
correlations. This enhanced response is consistent with growing
orientational organization below the
liquid-hexatic coexistence interval,
$0.700\leq\phi\leq0.716$~\cite{Bernard2011Oct}, although the present
data do not extend into that interval.

\subsubsection{Stable fluid branch of 3D hard spheres}
\label{subsubsec:results_eq_fluid_3D}

\begin{figure}[!h]
\centering
\includegraphics[width=\linewidth]
{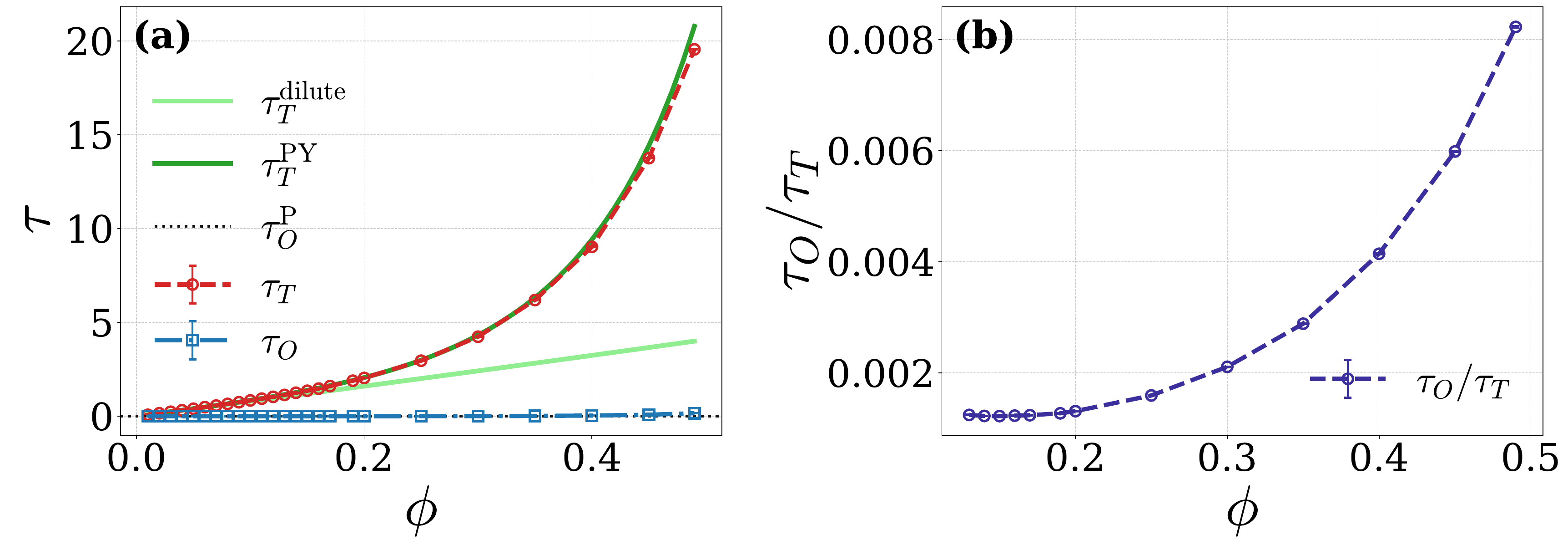}
\caption{Along the stable fluid branch of 3D equilibrium hard
spheres, (a) $\tau_T$ and $\tau_O$ as functions of $\phi$,
together with the dilute-limit prediction $\tau_T^{\rm dilute}=8\phi$
and the Percus-Yevick prediction for $\tau_T$, and (b) the ratio
$\tau_O/\tau_T$. The ratio remains below $1\%$ beyond the
Poisson-reference regime, showing that bond-orientational order
remains much smaller than translational order.}
\label{fig:eq_hs_fluid}
\end{figure}

For the 3D fluid, $\tau_T$ approaches the dilute-limit result at low
$\phi$, while the Percus-Yevick prediction~\cite{Torquato2002book}
provides an accurate reference over a broader range
[Fig.~\ref{fig:eq_hs_fluid}(a)]. The ratio $\tau_O/\tau_T$ increases
from approximately $1.3\times10^{-3}$ at $\phi=0.20$ to
$8.2\times10^{-3}$ at $\phi=0.49$, remaining below $1\%$ throughout the sampled range beyond the
Poisson-reference regime [Fig.~\ref{fig:eq_hs_fluid}(b)]. The enhancement observed in the 2D fluid is therefore absent in three dimensions: translational and bond-orientational correlations remain short-ranged, and $\tau_O$ remains very small relative to $\tau_T$. Because the ratio exhibits no comparably pronounced
increase, we do not apply the fractional-response analysis to this branch.

\subsubsection{Crystal branches}
\label{subsubsec:results_crystal_branches}

In the crystal branches, we use the radial cutoff-dependent definitions
in Eqs.~\eqref{eq:tauT_Kc} and \eqref{eq:tauO_Kc}, and both metrics
depend on the common radial wave-number cutoff $K_c$. To simplify
notation, throughout this subsection $\tau_T(K_c)$ and $\tau_O(K_c)$
denote $\tau_T(K_c;L)$ and $\tau_O(K_c;L)$, respectively, at the fixed
system size $L$ used for each crystal branch. Increasing $K_c$ includes
additional reciprocal vectors and raises the absolute values of
$\tau_T$ and $\tau_O$. Their ratio is much less cutoff sensitive and
can be compared with the perfect-crystal result
$\tau_O(K_c)/\tau_T(K_c)=1$ from
Eq.~\eqref{eq:tau_equal_perfect_crystal}.

\begin{figure}[!h]
\centering
\includegraphics[width=\linewidth]
{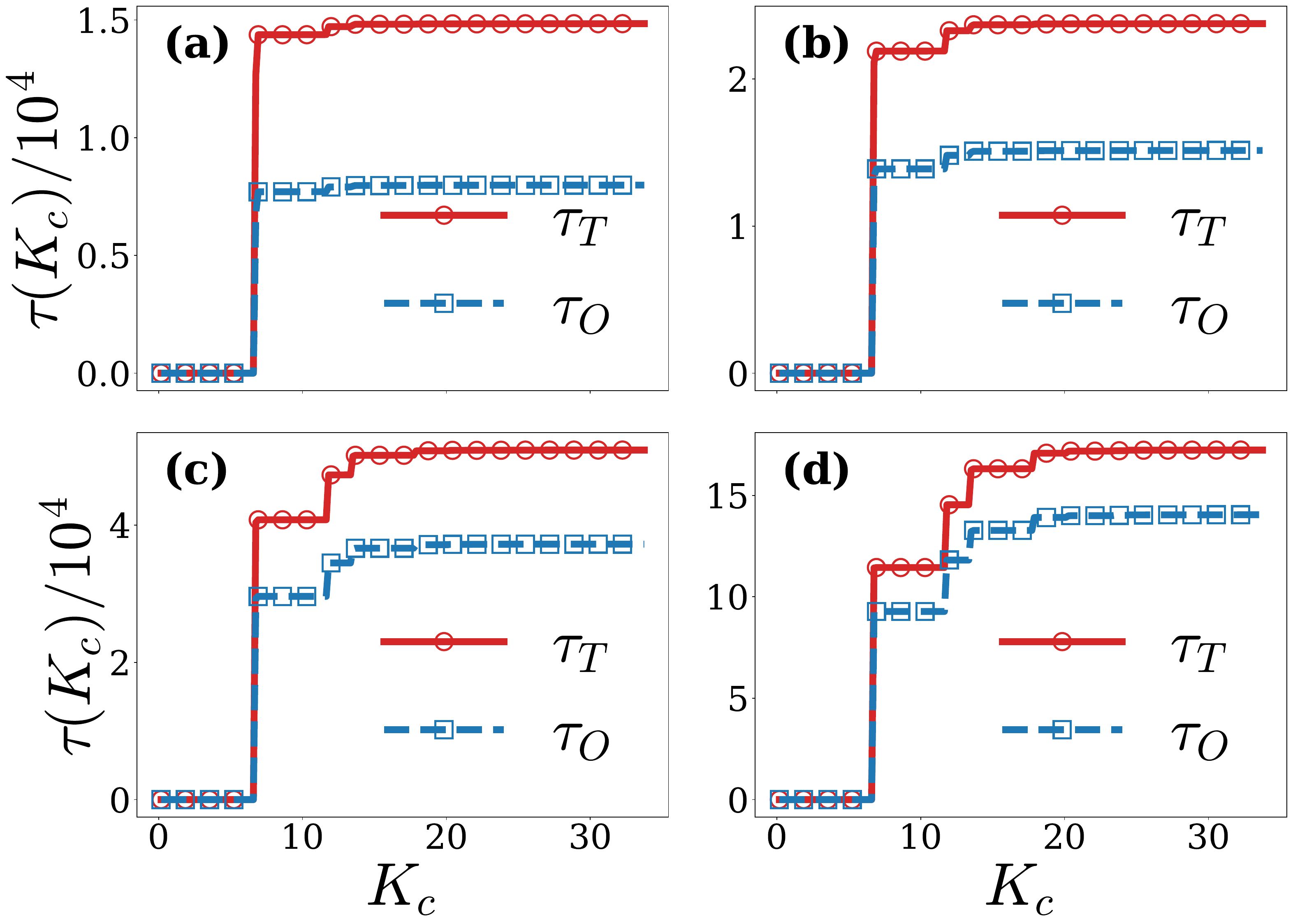}
\includegraphics[width=\linewidth]
{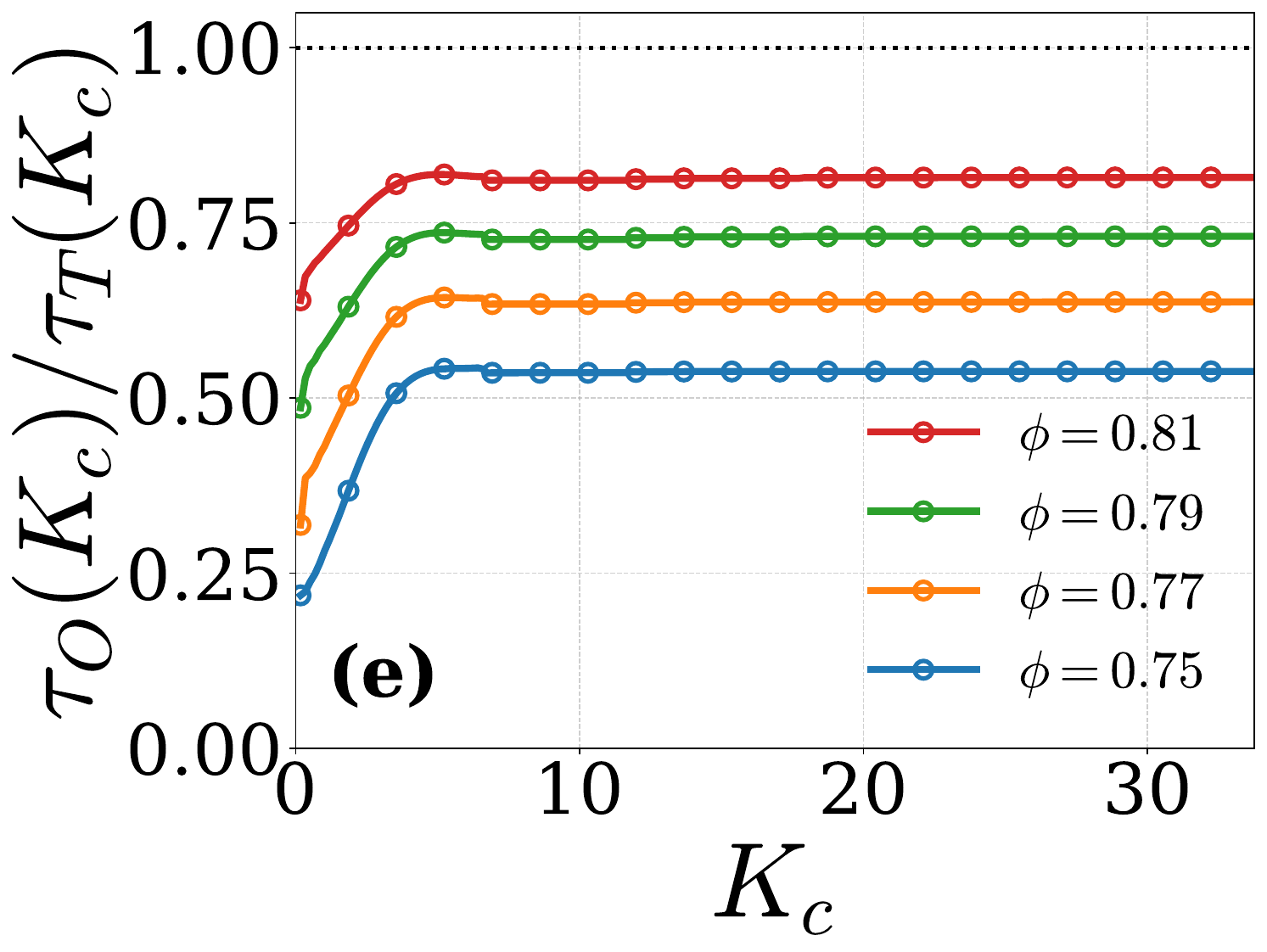}
\caption{The dependence of the order metrics on the cutoff wave number $K_c$ for
fixed system size $L$ along the 2D hard-disk crystal branch.
Panels (a)--(d) show $\tau_T/10^4$ and $\tau_O/10^4$ as functions of
the reciprocal-space cutoff $K_c$ for $\phi=0.75$, $0.77$, $0.79$,
and $0.81$, respectively. The stepwise increases occur as successive
Bragg shells enter the integration domain. Panel (e) shows
$\tau_O/\tau_T$ as a function of $K_c$ for all four packing fractions.
Although the absolute metrics receive substantial cutoff-dependent
contributions from the Bragg peaks,
over the higher-$K_c$ range shown,
their ratio approaches a stable,
packing-fraction-dependent value that increases with $\phi$.}
\label{fig:eq_crystal_2D_cutoff}
\end{figure}

For the 2D hard-disk crystal, both metrics increase
monotonically with $\phi$ at fixed $K_c$
[Figs.~\ref{fig:eq_crystal_2D_cutoff}(a)--\ref{fig:eq_crystal_2D_cutoff}(d)]. To characterize the
cutoff dependence at specified Bragg-shell cutoffs, we compare
$K_c=6.752$, corresponding to the first Bragg shell, with
$K_c=31.058$, the smallest sampled cutoff that fully includes the first
ten Bragg shells. At $K_c=6.752$, $\tau_O/\tau_T$ increases from
approximately $0.536$ at $\phi=0.75$ to $0.811$ at $\phi=0.81$;
the corresponding values at $K_c=31.058$ are $0.538$ and $0.815$
[Fig.~\ref{fig:eq_crystal_2D_cutoff}(e)]. Thus, although the individual
metrics receive substantial contributions as successive Bragg shells
are included, the ratio changes by less than $0.7\%$ upon increasing
the cutoff from the first-shell value to a cutoff including the first
ten Bragg shells for all four packing fractions. Translational and
bond-orientational order are comparable throughout the sampled branch
and become more nearly equal as $\phi$ increases.

\begin{figure}[!h]
\centering
\includegraphics[width=\linewidth]
{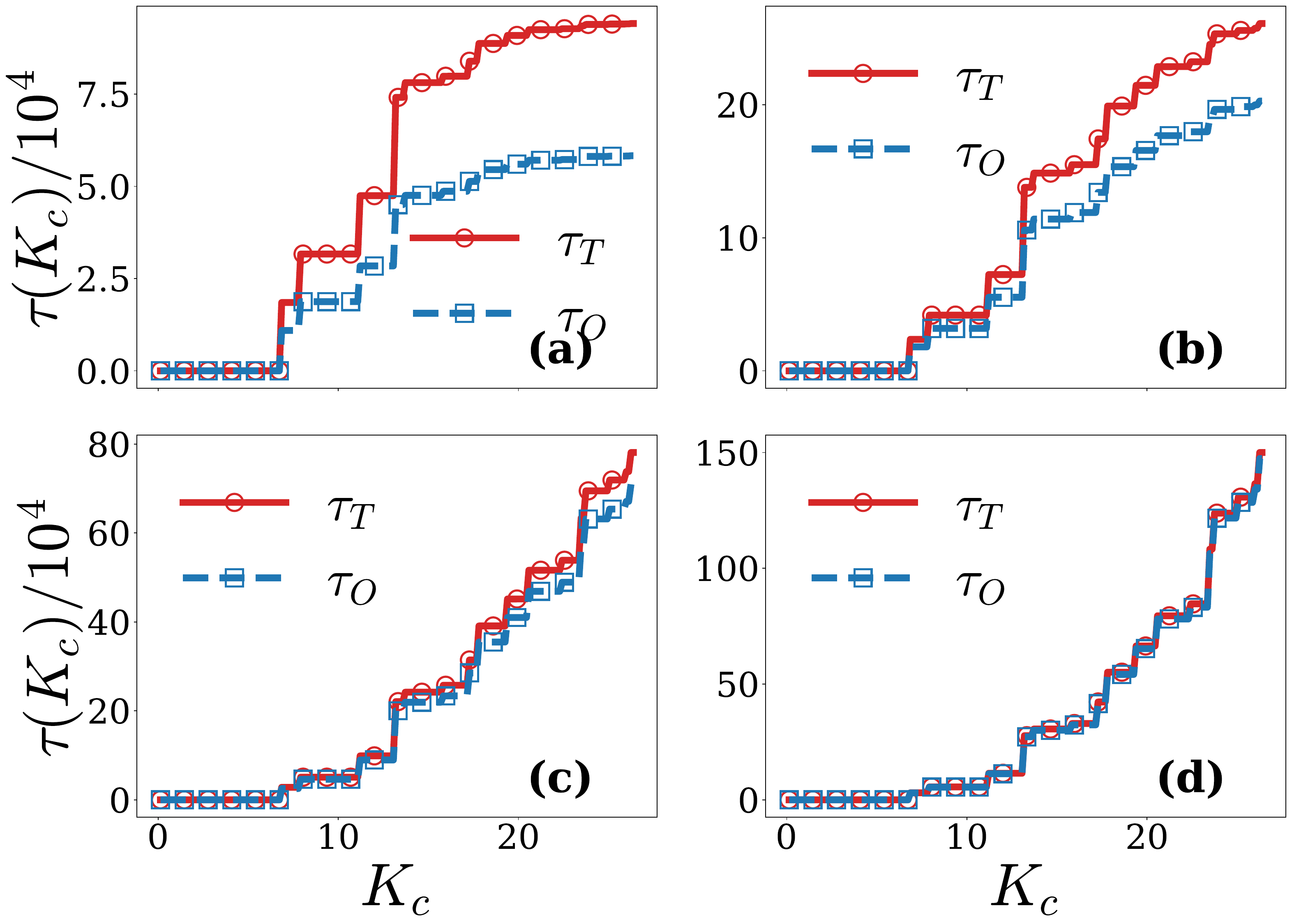}
\includegraphics[width=\linewidth]
{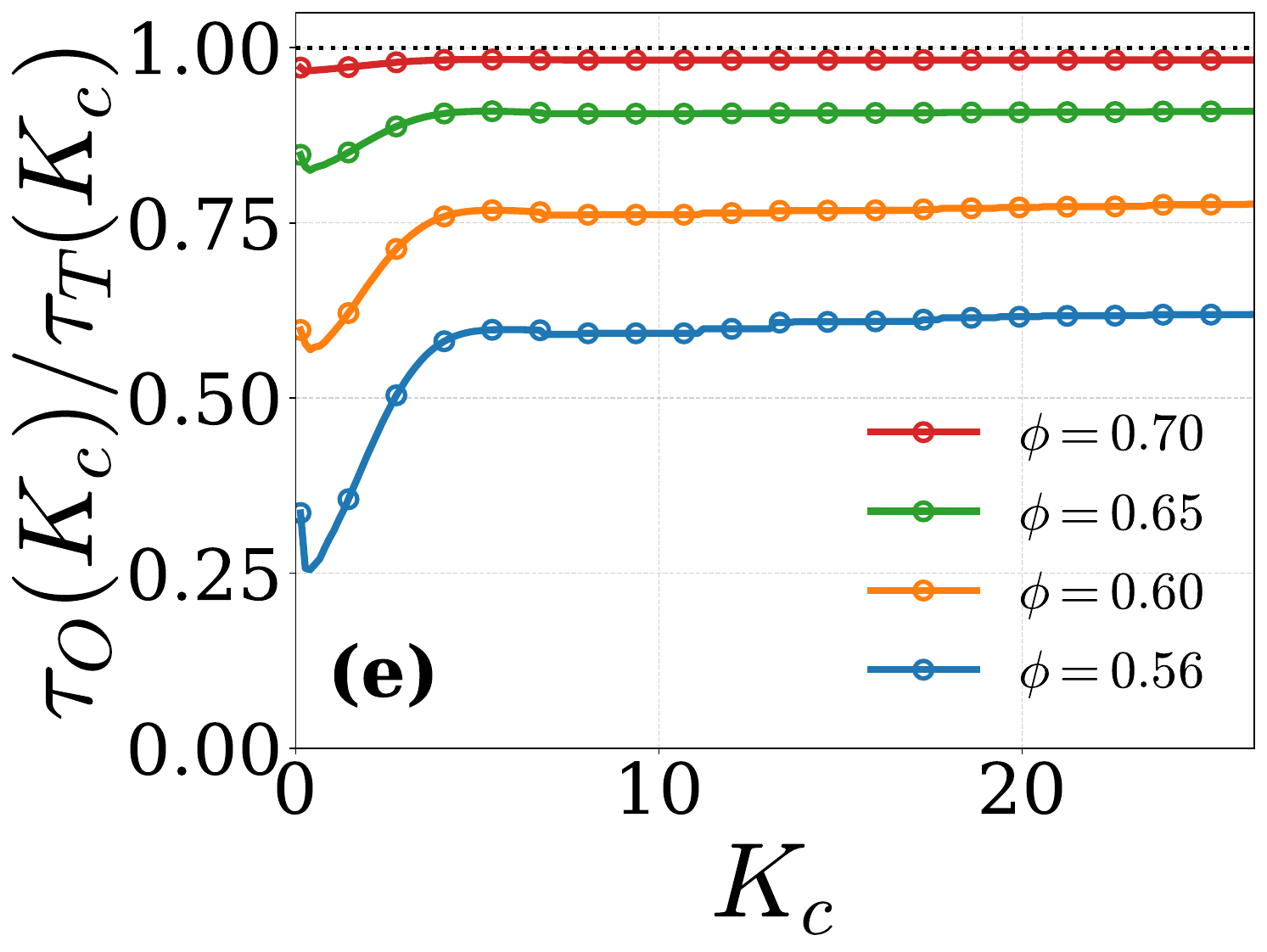}
\caption{The dependence of the order metrics on the cutoff wave number $K_c$ for
fixed system size $L$ along the 3D hard-sphere fcc crystal branch
using the 12 periodic nearest neighbors. Panels (a)--(d) show
$\tau_T/10^4$ and $\tau_O/10^4$ as functions of the reciprocal-space
cutoff $K_c$ for $\phi=0.56$, $0.60$, $0.65$, and $0.70$,
respectively. The stepwise increases occur as successive Bragg shells
enter the integration domain. Panel (e) shows $\tau_O/\tau_T$ as a
function of $K_c$ for all four packing fractions. Although the
individual metrics receive substantial cutoff-dependent contributions
from the Bragg peaks,
over the higher-$K_c$ range shown,
their ratio approaches a stable,
packing-fraction-dependent value that is nearly unity at the highest
packing fraction.}
\label{fig:eq_crystal_3D_12NN_cutoff}
\end{figure}

For the 3D crystal branch, the primary calculation
defines the orientational weights using the 12 periodic nearest
neighbors, corresponding to the physical first coordination shell of
the fcc lattice. With this definition, both metrics increase
monotonically with $\phi$ at fixed $K_c$
[Figs.~\ref{fig:eq_crystal_3D_12NN_cutoff}(a)--\ref{fig:eq_crystal_3D_12NN_cutoff}(d)]. To characterize the
cutoff dependence at specified Bragg-shell cutoffs, we consider the
smallest sampled cutoffs that fully include the first, fifth, tenth,
and fifteenth Bragg shells, $K_c=6.861$, $13.722$, $20.582$, and
$25.992$, respectively. At $\phi=0.56$, $\tau_O/\tau_T$ takes the
corresponding values $0.591$, $0.609$, $0.617$, and $0.619$, whereas
at $\phi=0.70$ the ratio remains essentially unchanged at approximately
$0.9825$--$0.9826$
[Fig.~\ref{fig:eq_crystal_3D_12NN_cutoff}(e)]. Thus, the cutoff
dependence of the ratio is most pronounced at the lowest packing
fraction, where it changes by approximately $4.8\%$ from the first to
the fifteenth Bragg shell. However, even at $\phi=0.56$, the change
from the tenth to the fifteenth shell is only approximately $0.3\%$,
showing that the ratio becomes progressively less sensitive to the
inclusion of higher Bragg shells. At $\phi=0.70$, the corresponding
first-to-fifteenth-shell change is only approximately $0.01\%$.
A diagnostic comparison using periodic Delaunay neighbors is presented
in Appendix~\ref{app:3D_crystal_delaunay}. It leaves $\tau_T$
unchanged but reduces $\tau_O$ and $\tau_O/\tau_T$ because the
Delaunay construction includes neighbors beyond the first coordination
shell.

Taken together, the equilibrium results reveal a clear fluid-crystal
distinction. Bond-orientational order remains subdominant along both
fluid branches, although its relative growth becomes pronounced near
the upper end of the sampled 2D branch. By contrast, for the neighbor
definitions used in the primary calculations, $\tau_T$ and $\tau_O$
are comparable along both crystal branches and approach equality as
the perfect-crystal limit is approached. The 2D crystal values are
based on one configuration at each packing fraction, whereas the 3D
values are averages over ten configurations.

\subsection{RSA packings}
\label{subsec:results_rsa}

\begin{figure}[!h]
\centering
\includegraphics[width=\linewidth]
{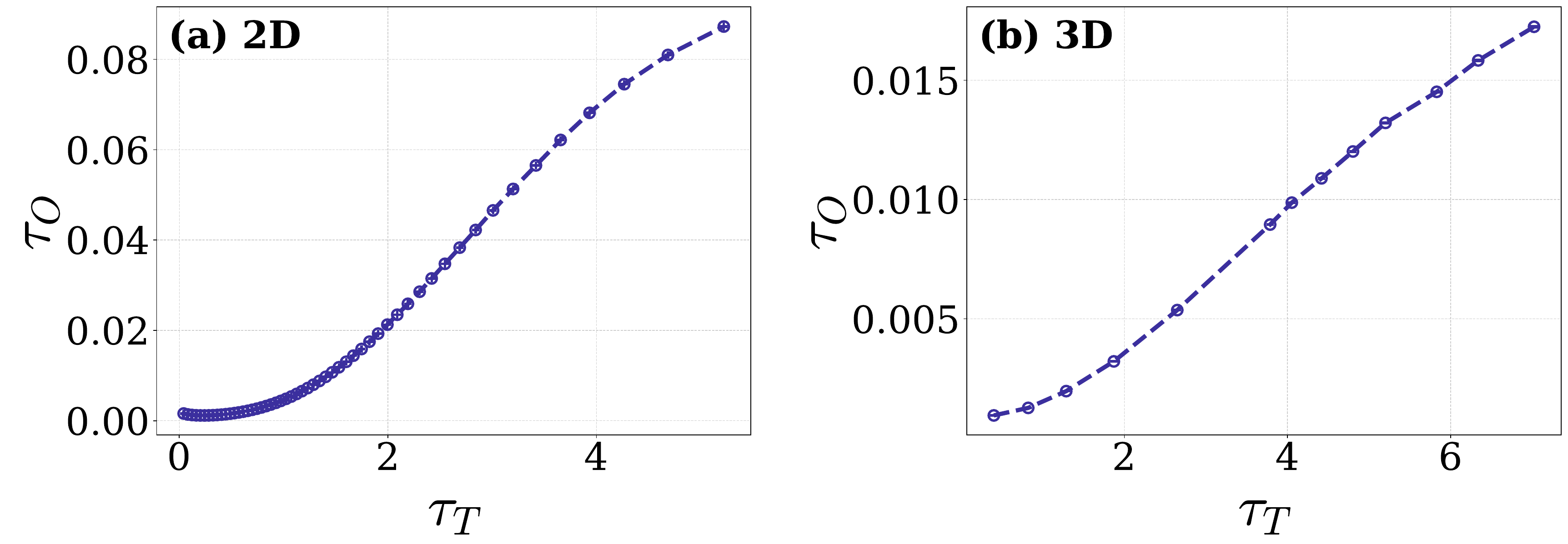}
\caption{Parametric relation between $\tau_O$ and $\tau_T$ along
the RSA branches of (a) 2D hard disks and (b) 3D hard spheres.}
\label{fig:rsa_parametric}
\end{figure}

\begin{figure}[!h]
\centering
\includegraphics[width=\linewidth]
{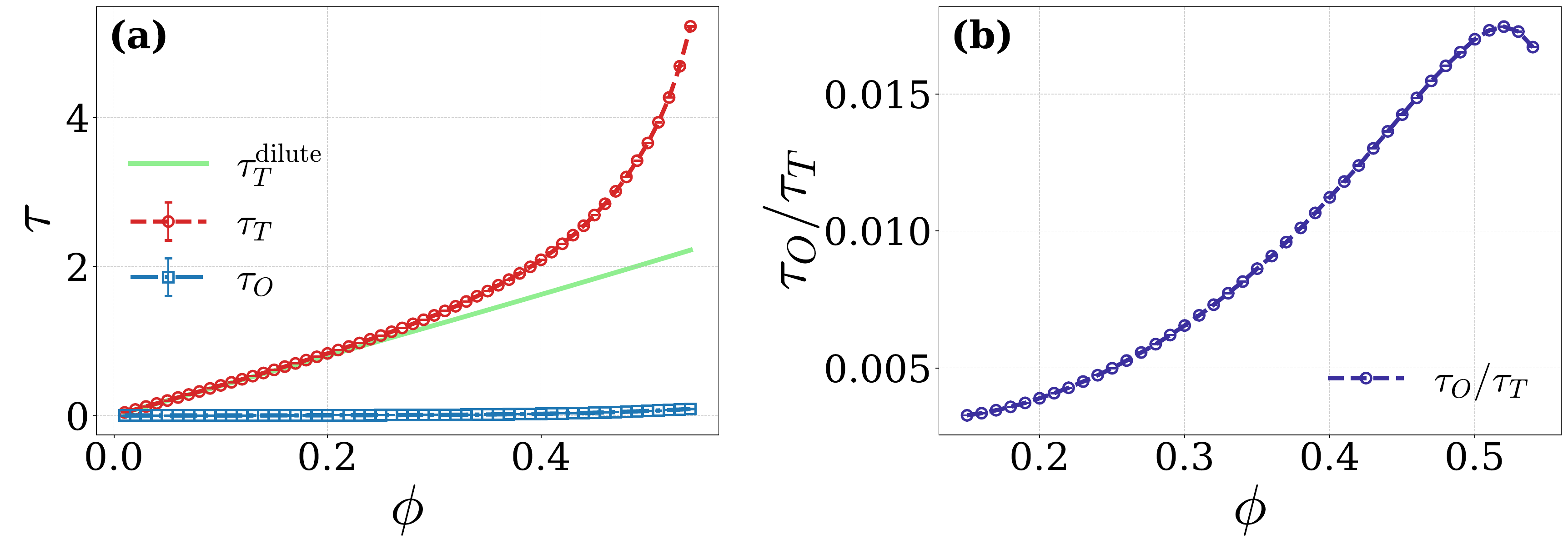}
\caption{Along the 2D RSA branch, (a) $\tau_T$ and $\tau_O$ as
functions of $\phi$, together with the dilute-limit prediction
$\tau_T^{\rm dilute}=4\phi$ for $\tau_T$, and (b) the ratio
$\tau_O/\tau_T$. Although $\tau_O$ remains subdominant, its magnitude
relative to $\tau_T$ increases appreciably over most of the branch
before decreasing slightly near saturation because of the more rapid
growth of $\tau_T$ compared with $\tau_O$.}
\label{fig:rsa_2d}
\end{figure}

\begin{figure}[!h]
\centering
\includegraphics[width=\linewidth]
{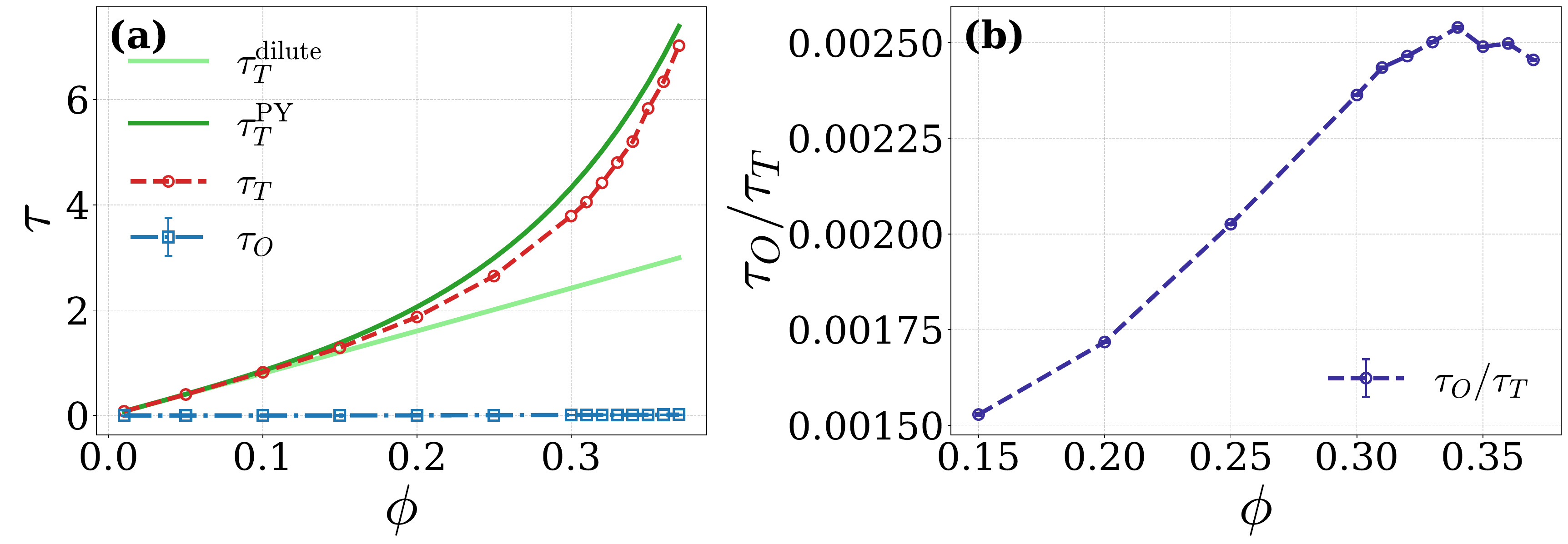}
\caption{Along the 3D RSA branch,
(a) $\tau_T$ and $\tau_O$ as functions of $\phi$, together with
the dilute-limit prediction $\tau_T^{\rm dilute}=8\phi$ and the
Percus-Yevick prediction for $\tau_T$, and (b) the ratio
$\tau_O/\tau_T$. The ratio remains much smaller than unity throughout
the branch and exhibits a slight downturn near saturation, as in two
dimensions, reflecting the more rapid growth of $\tau_T$ compared with
$\tau_O$.}
\label{fig:rsa_3d}
\end{figure}

Figure~\ref{fig:rsa_parametric} shows the parametric trajectories of the
2D and 3D RSA systems from the dilute regime to just below saturation.
In both dimensions, $\tau_O$ increases monotonically with $\tau_T$, but
bond-orientational order remains a small fraction of translational
order. Unless otherwise stated, the comparisons
of the order metrics refer to sampled states beyond the
Poisson-reference regime.

At low packing fractions, $\tau_T$ follows its dilute-limit prediction
in both dimensions [Figs.~\ref{fig:rsa_2d}(a)
and~\ref{fig:rsa_3d}(a)]. The Percus-Yevick curve in
Fig.~\ref{fig:rsa_3d}(a) is included only as an equilibrium hard-sphere
reference in three dimensions. In two dimensions $\tau_O/\tau_T$ increases from approximately
$3.89\times10^{-3}$ at $\phi=0.20$ to $1.67\times10^{-2}$ at
$\phi=0.54$, attaining a maximum of approximately $1.75\times10^{-2}$ at
$\phi=0.52$ [Fig.~\ref{fig:rsa_2d}(b)]. In three dimensions the variation is smaller:
the ratio increases from approximately $1.72\times10^{-3}$ at
$\phi=0.20$ to $2.45\times10^{-3}$ at $\phi=0.37$, with a maximum of
approximately $2.54\times10^{-3}$ at $\phi=0.34$
[Fig.~\ref{fig:rsa_3d}(b)]. Hence, $\tau_O$ remains below $2\%$ of
$\tau_T$ in two dimensions and below $0.3\%$ in three.
The ratio turns slightly downward close to saturation in both dimensions
even though $\tau_O$ itself continues to increase.
This downturn reflects the more rapid growth of
$\tau_T$ compared with $\tau_O$ near saturation.
It therefore represents a change in the relative growth rates of the
two metrics, not a loss of bond-orientational order.

\subsection{Comparison of equilibrium and RSA packings}
\label{subsec:comparison_equilibrium_RSA}

\begin{figure}[!h]
\centering
\includegraphics[width=0.49\linewidth]
{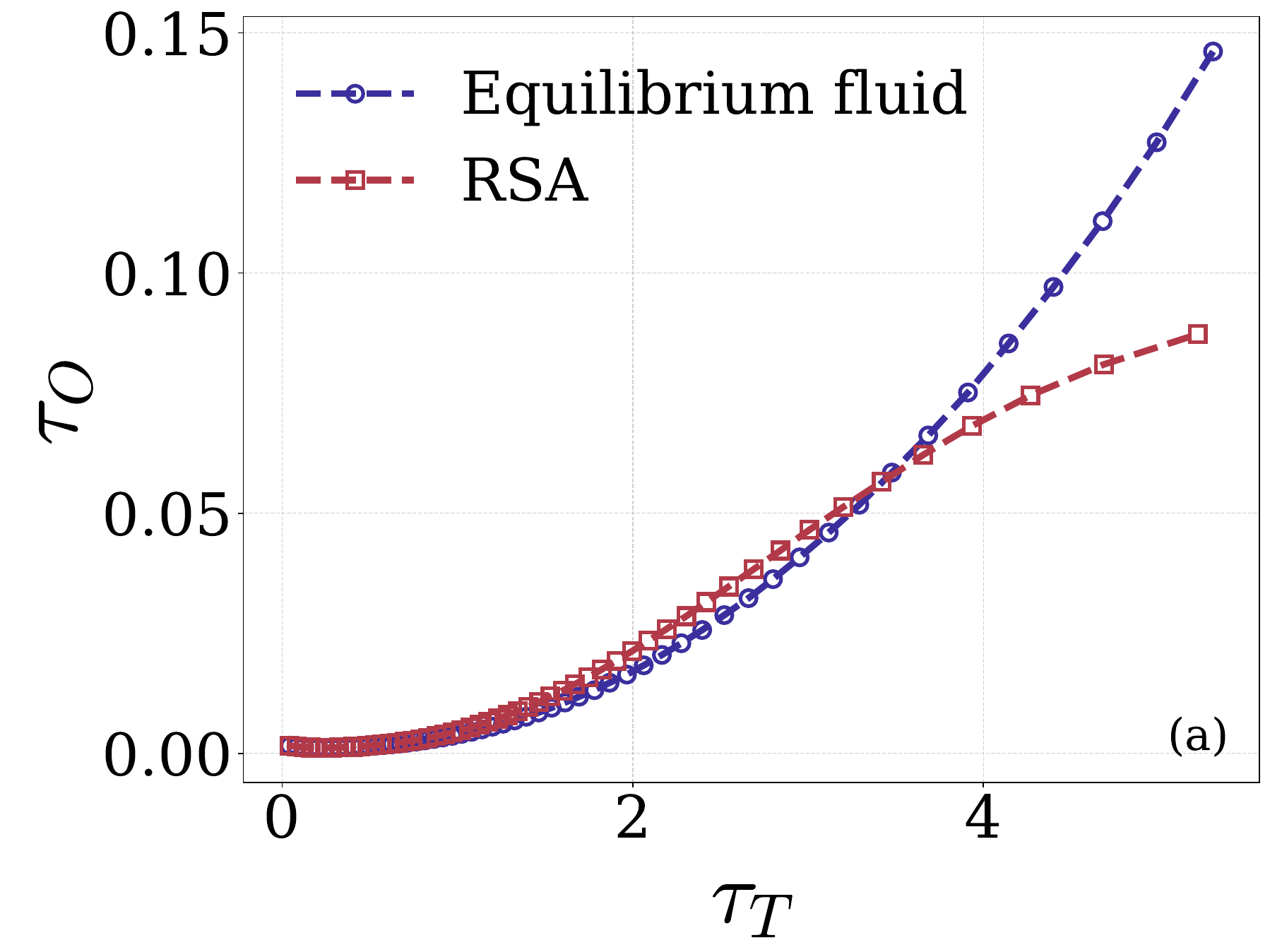}
\includegraphics[width=0.49\linewidth]
{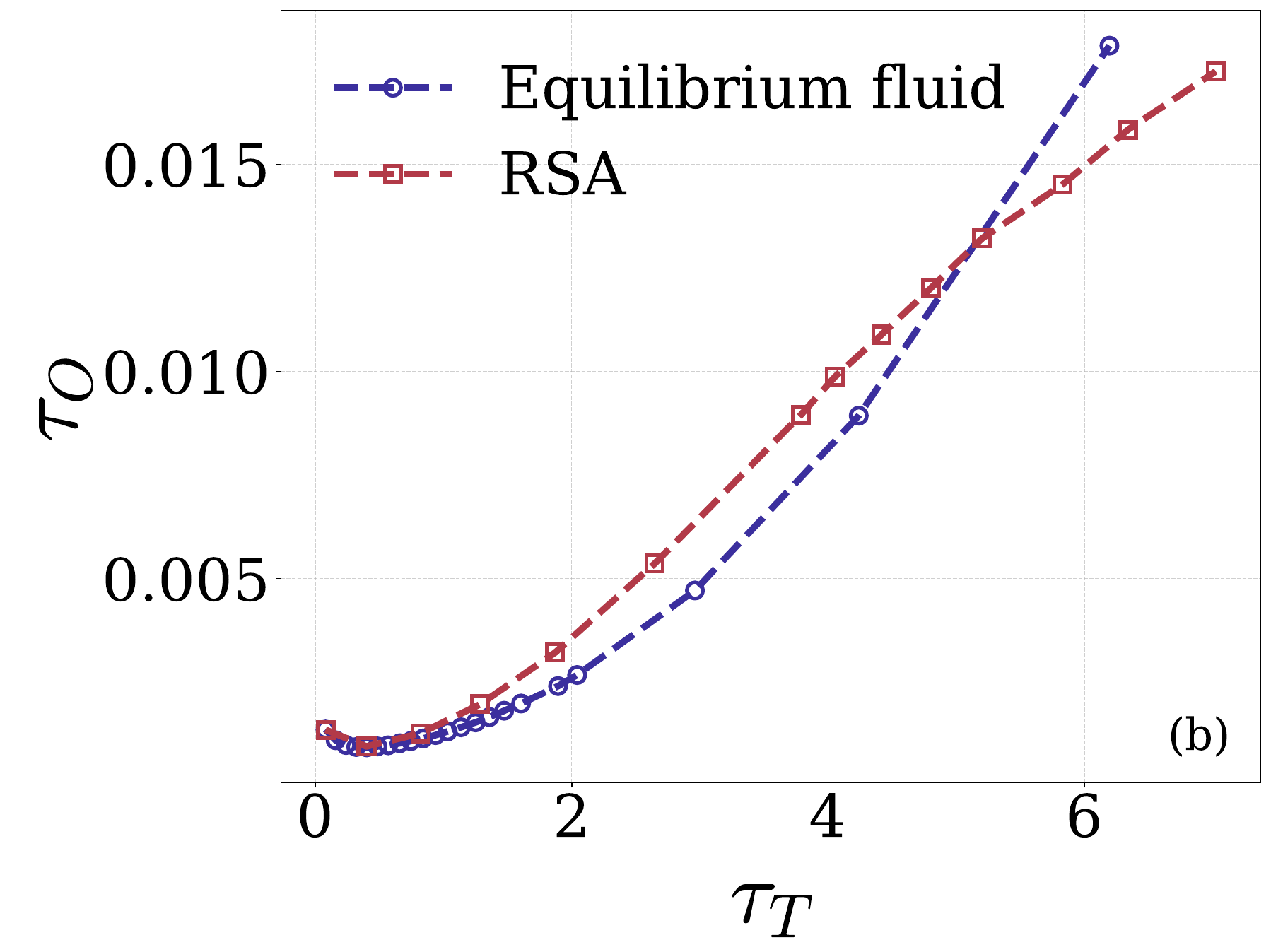}
\caption{Parametric comparison of $\tau_O$ and $\tau_T$ for
equilibrium fluids and RSA packings in (a) two dimensions and (b) three dimensions. The two
preparation protocols follow distinct trajectories in the
$(\tau_T,\tau_O)$ plane.}
\label{fig:tauO_vs_tauT_equilibrium_RSA}
\end{figure}

Figure~\ref{fig:tauO_vs_tauT_equilibrium_RSA}
compares the equilibrium-fluid and RSA trajectories in the
$(\tau_T,\tau_O)$ plane. In both dimensions, RSA packings have larger
$\tau_O$ at moderate $\tau_T$, whereas the equilibrium-fluid branch has
larger $\tau_O$ at sufficiently high $\tau_T$. Thus, comparable $\tau_T$
can correspond to different $\tau_O$ depending on the preparation
protocol.
This distinction is most pronounced in two dimensions, where the equilibrium-fluid
trajectory bends upward at large $\tau_T$ because the relative growth of
$\tau_O$ exceeds that of $\tau_T$ near the upper end of the sampled
branch, whereas the RSA trajectory shows no analogous enhancement. In three dimensions $\tau_O$ remains much smaller than $\tau_T$ for both protocols, but
their trajectories in the $(\tau_T,\tau_O)$ plane remain distinct.

\subsection{Stealthy hyperuniform ground states}
\label{subsec:results_shu}

\begin{figure}[!h]
\centering
\includegraphics[width=0.62\linewidth]
{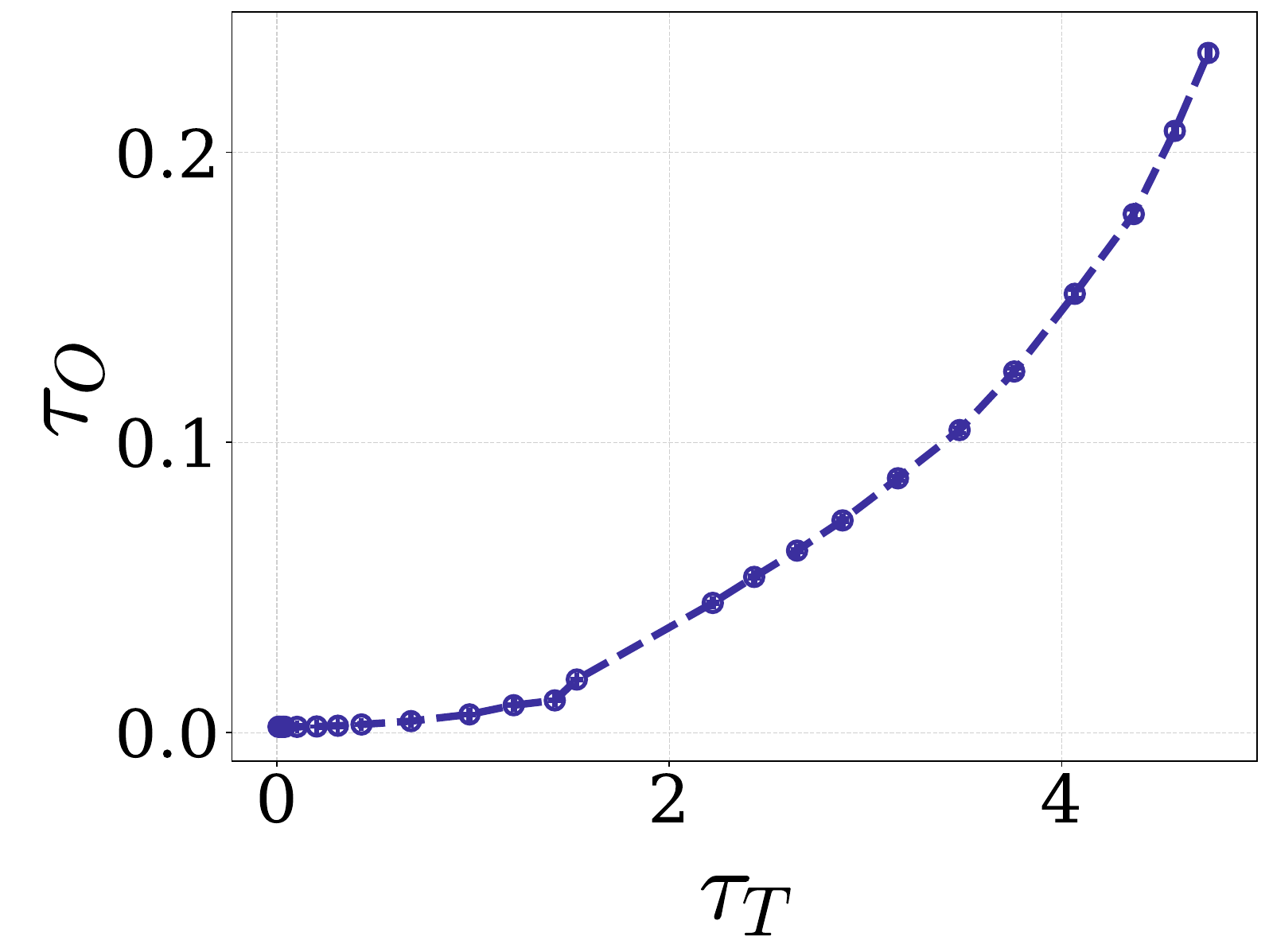}
\caption{Parametric relation between $\tau_O$ and $\tau_T$
within the 2D disordered SHU phase.}
\label{fig:tauO_vs_tauT_SHU}
\end{figure}

\begin{figure}[!h]
\centering
\includegraphics[width=\linewidth]
{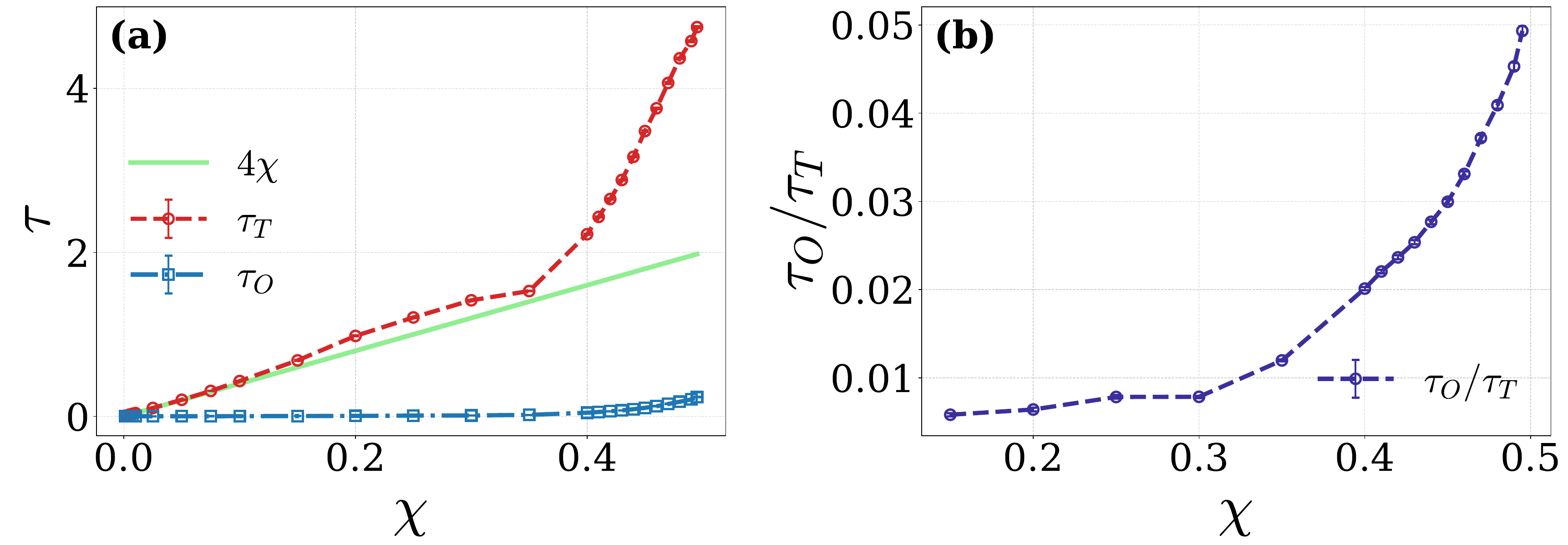}
\caption{Within the 2D disordered SHU phase,
(a) $\tau_T$ and $\tau_O$ as functions of $\chi$, together with
the small-$\chi$ prediction $\tau_T\simeq4\chi$, and
(b) the ratio $\tau_O/\tau_T$. Although $\tau_O$ remains
subdominant, its magnitude relative to $\tau_T$ increases as the
ordering threshold at $\chi=1/2$ is approached.}
\label{fig:tau_levels_ratio_SHU}
\end{figure}

\begin{figure}[!h]
\centering
\includegraphics[width=\linewidth]
{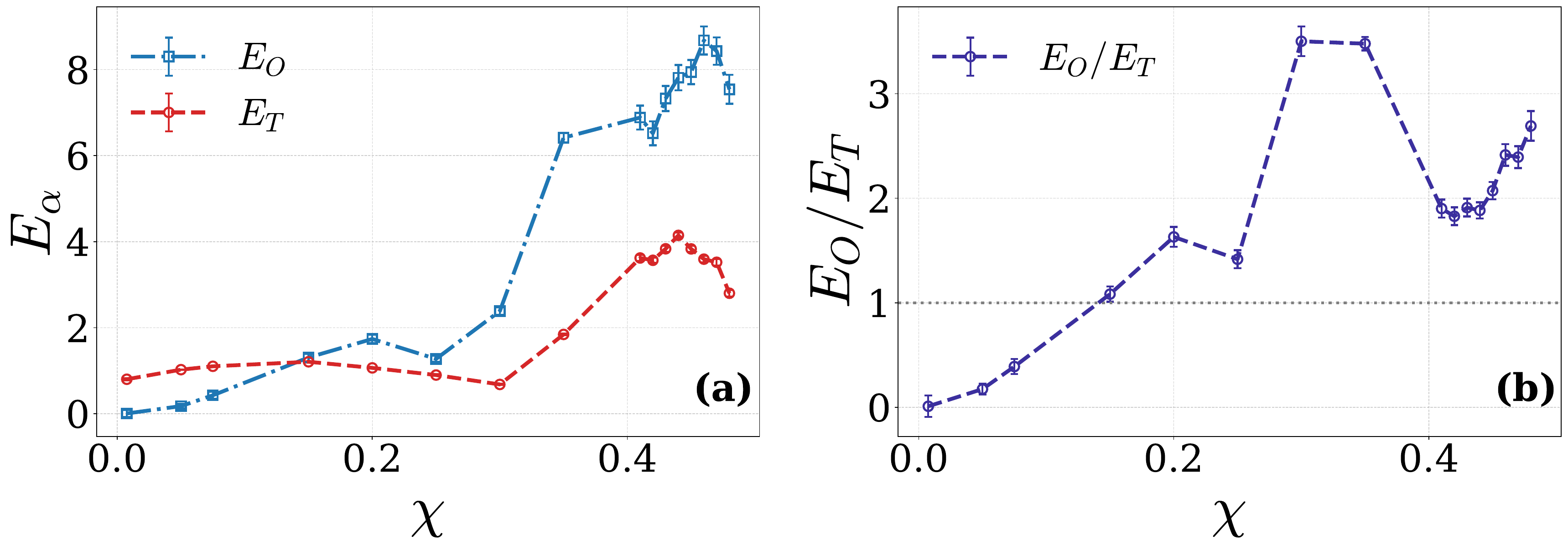}
\caption{Sensitivities $E_T$ and $E_O$ and their ratio
$E_O/E_T$ within the 2D disordered SHU phase. The responses vary
across different ranges of $\chi$, while near the ordering
threshold the fractional growth of $\tau_O$ exceeds that of
$\tau_T$.}
\label{fig:elasticity_SHU}
\end{figure}

We finally consider the disordered 2D SHU ground states specified in
Sec.~\ref{subsec:shu}, using $\chi$ as the control parameter.
Hyperuniformity is determined by the behavior of $S(k)$ as $k\to0$,
whereas $\tau_T$ integrates contributions over the full wave-number range
[Eq.~\eqref{eq:tauT_iso_fourier}]. The two quantities therefore describe different
aspects of the structure. Within the sampled disordered SHU phase, $\tau_O$
increases monotonically with $\tau_T$
(Fig.~\ref{fig:tauO_vs_tauT_SHU}).
The translational metric increases throughout the sampled range and
follows the small-$\chi$ prediction $\tau_T\simeq4\chi$
[Fig.~\ref{fig:tau_levels_ratio_SHU}(a)]. Although orientational order
remains subdominant, $\tau_O/\tau_T$ rises from approximately
$6.42\times10^{-3}$ at $\chi=0.20$ to
$4.93\times10^{-2}$ at
$\chi=0.495$ [Fig.~\ref{fig:tau_levels_ratio_SHU}(b)]. Near the nominal
disorder-to-order threshold at $\chi=1/2$, $E_O>E_T$, and $E_O/E_T$
increases overall from approximately $1.90$ at $\chi=0.41$ to $2.69$ at
$\chi=0.48$ [Fig.~\ref{fig:elasticity_SHU}]. Thus, $\tau_O/\tau_T$ grows as the threshold is approached, even though $\tau_O$ remains much smaller than $\tau_T$.

\section{CONCLUSIONS AND DISCUSSION}
\label{sec:conclusions}

The principal methodological advance of this work is that it places
translational and bond-orientational order on a common two-point
statistical footing. Whereas conventional metrics often quantify either positional order or spatially averaged local orientational order, the pair $(\tau_T,\tau_O)$, defined in
Eqs.~\eqref{eq:tauT} and \eqref{eq:tauO}, measures the integrated squared
magnitude of the corresponding correlations and is therefore sensitive to both their amplitude and spatial persistence. Across nonhyperuniform equilibrium
fluids, equilibrium crystals, nonhyperuniform RSA packings, and disordered
SHU ground states, the pair $(\tau_T,\tau_O)$ reveals structural
distinctions associated with preparation protocol, dimensionality, and
proximity to ordering.

Across the systems studied, $\tau_T$ and $\tau_O$ increase monotonically
together along the sampled control-parameter paths (i.e., packing fraction
$\phi$ in the case of the packing models and the stealthiness parameter $\chi$ in
the case of disordered stealthy hyperuniform ground states), but their
relative magnitudes vary substantially.
Bond-orientational order remains subdominant along the equilibrium fluid
branches of the hard-particle systems, the RSA branches, and
within the disordered SHU phase, whereas $\tau_T$ and $\tau_O$ become comparable
in magnitude along the sampled equilibrium crystal branches of the
hard-particle systems. The sensitivities further show that
$\tau_O$ grows more rapidly than $\tau_T$ in fractional terms near the
upper end of the 2D equilibrium fluid branch and near the SHU ordering
threshold. 

It is useful to comment on the effect of dimensionality on the coupling between translational and orientational ordering in the case of the two hard-sphere models that we have considered.  In equilibrium sphere packings, we have shown that orientational ordering relative to translational ordering is stronger along the fluid branch in two dimensions than in three. This weakening of the ratio also holds qualitatively as we go from 2D to 3D RSA packings. Moreover, the coupling between translational and orientational ordering is comparable along the crystal branches in two and three dimensions. We note that both translational and orientational ordering become weaker along the fluid branch of equilibrium hard spheres as well as RSA hard spheres 
as the space dimension increases beyond three
because  hard-sphere structures decorrelate with increasing $d$ \cite{Stillinger2006Jan, skoge2006, Torquato2012Feb}.

Several qualifications concerning the interpretation of these metrics
should be noted. Because $\tau_T$ and $\tau_O$ are scalar integrated
measures, it is clear that equal values do not imply identical underlying
correlation functions or structures. The observed hierarchy
$\tau_O<\tau_T$ is specific to the systems sampled here and should not be
taken to be universal. In a hexatic phase, where bond-orientational
correlations are quasi-long-ranged while translational correlations remain
short-ranged~\cite{Nelson1979Mar,Torquato2026Mar}, one may instead have
$\tau_O>\tau_T$.
The local orientational weights entering $\tau_O$, $f_j=\psi_6(j)$ in
two dimensions [Eq.~\eqref{eq:psi6}] and
$\vf_j=\{Q_{6m}(j)\}_{m=-6}^{6}$ in three [Eqs.~\eqref{eq:f3D} and~\eqref{eq:Q6m}], depend on the definition of
neighboring particles and bonds~\cite{vanMeel2012Jun, Mickel2013Jan}. Comparisons of $\tau_O$ must therefore
employ a consistent and physically appropriate neighbor-construction
protocol. Determining how the relation between $\tau_O$ and $\tau_T$
depends on this choice, particularly for disordered systems, remains an
interesting direction for future study. 
In the crystalline phases in both two and three dimensions, neither
metric has a finite thermodynamic limit as $L\to\infty$. At fixed
system size $L$, however, $\tau_T(K_c;L)$ and $\tau_O(K_c;L)$, defined
in Eqs.~\eqref{eq:tauT_Kc} and~\eqref{eq:tauO_Kc}, respectively,
approach system-size-dependent values as $K_c\to\infty$. Over the range
of cutoffs considered, the ratio $\tau_O/\tau_T$ is substantially less
cutoff sensitive than the individual metrics, and its stability has
been checked directly as a function of $K_c$.

In future condensed-matter and statistical-physics studies, these metrics
could be used to map structural pathways through freezing, melting, glass
formation, jamming, self-assembly, and driven nonequilibrium
transitions~\cite{Steinhardt1983Jul,Truskett2000Jul,
Torquato2009Inverse,Digregorio2018Active}. The joint $(\tau_T,\tau_O)$ representation may help distinguish
phases or metastable states with similar conventional pair structure,
identify growing orientational organization before
crystallization~\cite{Russo2012Crystallization}, and compare structures
generated by different dynamics or processing
histories~\cite{Truskett2000Jul}. Extensions to other rotational
symmetries, multicomponent systems, anisotropic particles, quasicrystals,
active matter, and spin- or director-weighted configurations could
provide symmetry-specific order
coordinates~\cite{Steinhardt1983Jul,Torquato2009Inverse,
Digregorio2018Active}. A further direction is to construct analogous
order metrics directly from explicit higher-order
correlation functions, thereby quantifying structural information not
fully captured by the present pair $(\tau_T,\tau_O)$. The metric pair could also serve as collective
variables in enhanced sampling~\cite{Auer2001Nucleation}, data-driven
phase classification~\cite{Carrasquilla2017Machine}, and inverse-design
optimization~\cite{Torquato2009Inverse,Torquato2022LocalOrder,
Maher2024Sep,Shi2025Construction}, especially if future work
establishes how particular regions of the $(\tau_T,\tau_O)$ plane relate
to thermodynamic stability, kinetics, transport, mechanical response, or
optical properties~\cite{Kim2024OpticalTransport,Shi2025Construction}.

\begin{acknowledgments}
We thank Samuel Dawley for numerical assistance with the independent
verification of the equilibration of the crystal-branch configurations. We also thank Paul Steinhardt for valuable discussions. A.M. and S.T. were supported by the
U.S. Army Research Office under Cooperative Agreement No. W911NF-22-2-0103,
and we acknowledge the Princeton Institute for Computational Science and
Engineering for the computational resources used in this work.
\end{acknowledgments}

\section*{Data availability}
There are no publicly available research data or software supporting this paper.
Requests for further information or data should be sent to the authors.

\begin{appendix}

\section{Representative correlation functions and structure factors}
\label{app:representative_correlations}

\begin{figure}[!h]
\centering
\includegraphics[width=\linewidth]
{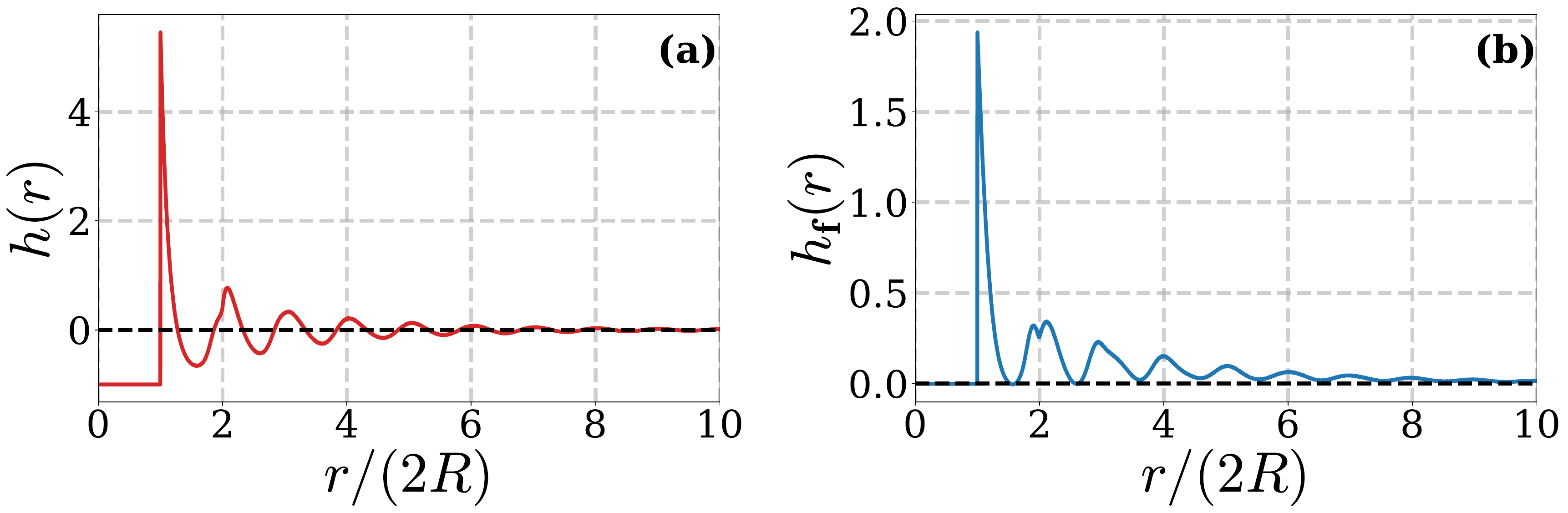}
\includegraphics[width=\linewidth]
{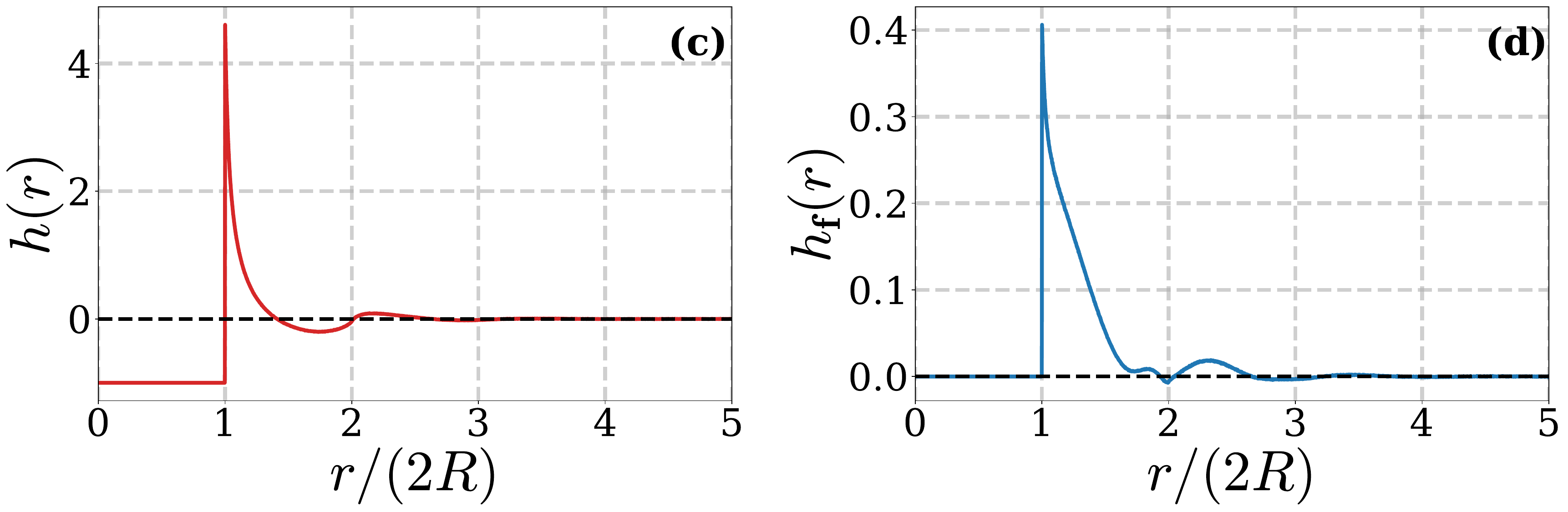}
\caption{Correlation functions for the 2D equilibrium hard-disk fluid
at $\phi=0.68$, the sampled fluid state closest to the liquid-hexatic
transition (a), (b) and 2D RSA disk packings at $\phi=0.54$, near
saturation (c), (d). Panels (a) and (c) show $h(r)$, and panels (b) and (d) show $h_{\vf}(r)$, defined in Eqs.~\eqref{eq:hr} and \eqref{eq:hf},
respectively. The orientational weight $f_j=\psi_6(j)$ is defined in
Eq.~\eqref{eq:psi6}. Separations are expressed as $r/(2R)$, where $R$
is the disk radius. In both systems, the correlations decay well within
the size of the simulation boxes.}
\label{fig:correlations_2D_liquid_RSA}
\end{figure}

\begin{figure}[!h]
\centering
\includegraphics[width=\linewidth]
{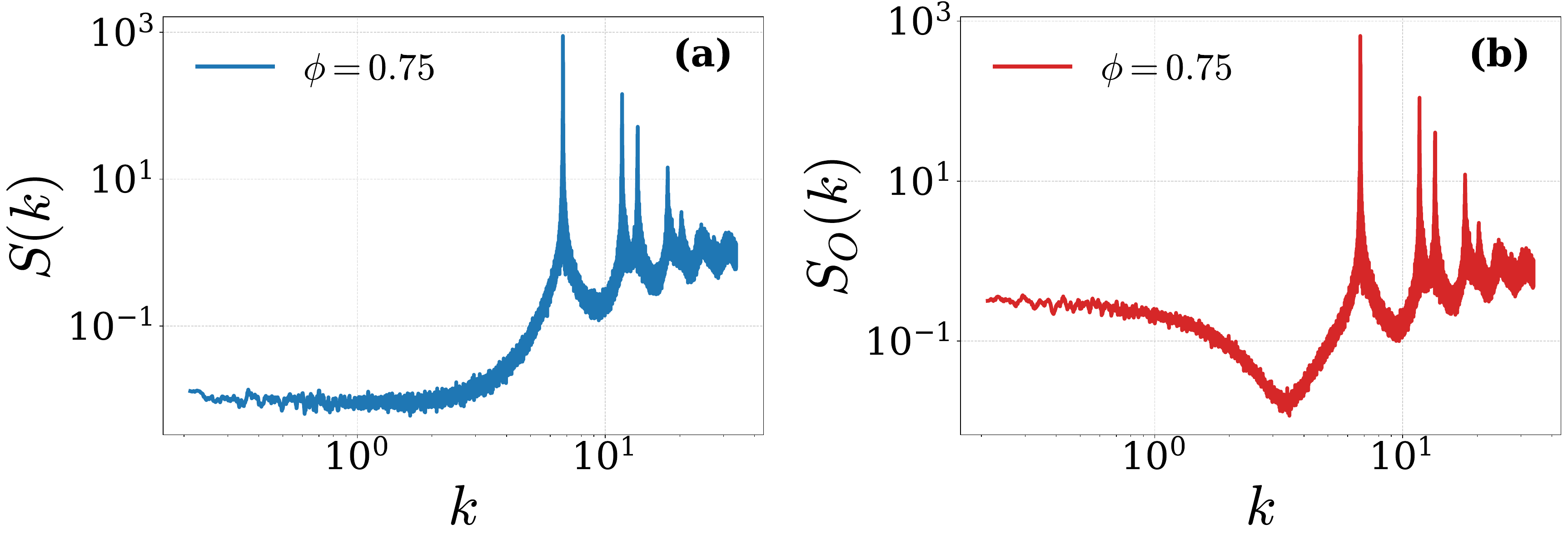}
\caption{Radially averaged structure factors for the 2D equilibrium
hard-disk crystal at $\phi=0.75$: (a) the translational structure
factor $S(k)$ and (b) the bond-orientational structure factor $S_O(k)$.
The orientational weight $f_j=\psi_6(j)$ is defined in
Eq.~\eqref{eq:psi6}.}
\label{fig:structure_factors_2D_crystal_phi075}
\end{figure}

\begin{figure}[!htb]
\centering
\includegraphics[width=\linewidth]
{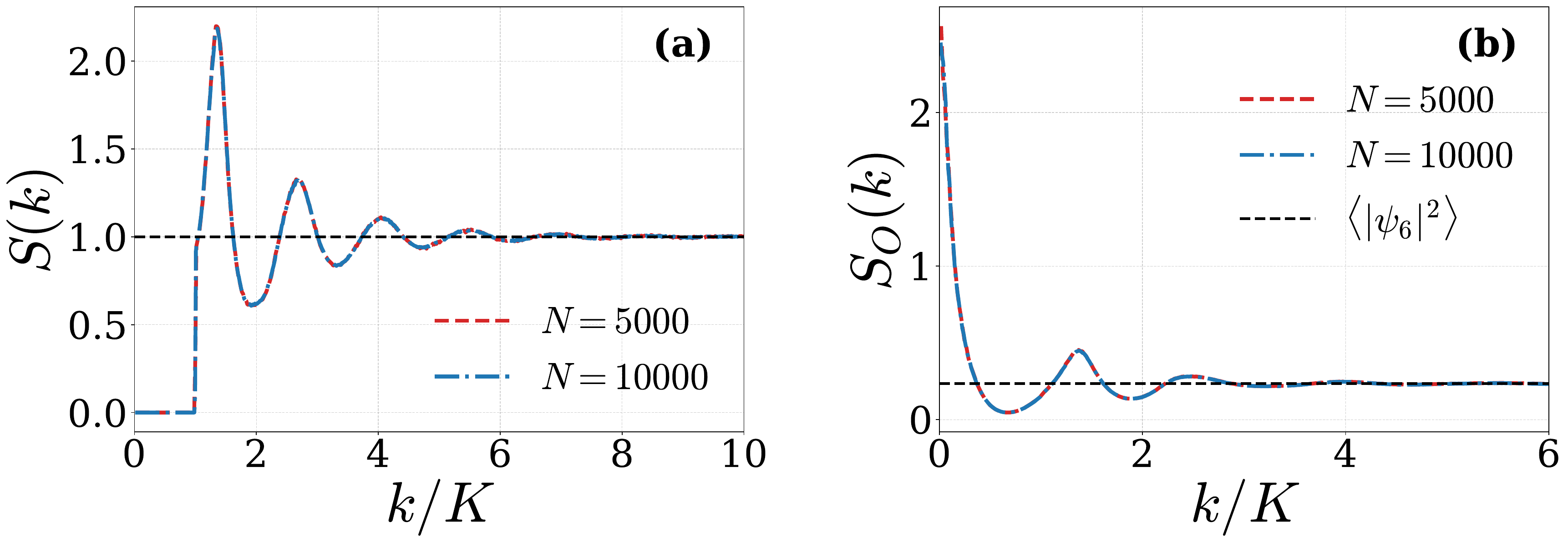}
\caption{Representative radially averaged structure factors for the 2D
disordered SHU ground-state ensemble at $\chi=0.495$, the sampled state
closest to the ordering threshold at $\chi=1/2$: (a) the unweighted
structure factor $S(k)$ and (b) the bond-orientational structure factor
$S_O(k)$. The wave number is reported as the dimensionless quantity
$k/K$, where $K$ is the width of the stealthy exclusion region. Both structure factors are essentially unchanged upon
increasing the system size from $N=5000$ to $N=10000$. The horizontal black dashed lines indicate unity in
panel (a) and the self-contribution $\langle|\psi_6|^2\rangle$ in
panel (b).}
\label{fig:structure_factors_SHU_chi0495}
\end{figure}

\begin{figure}[!htb]
\centering
\includegraphics[width=\linewidth]
{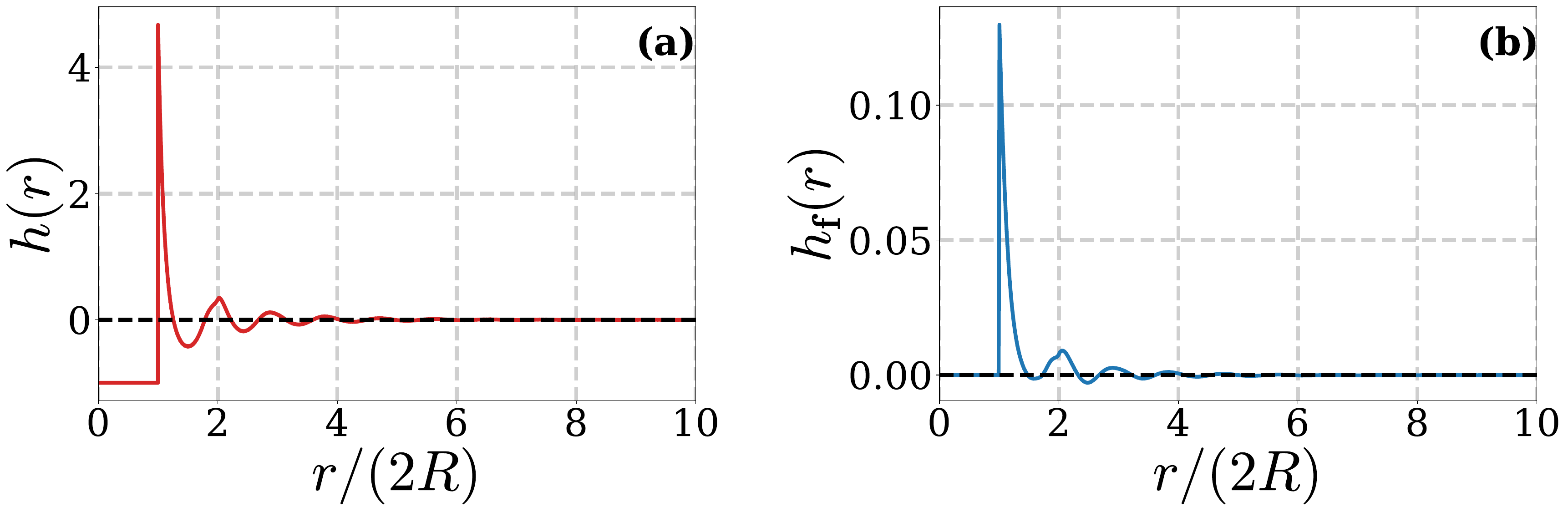}
\includegraphics[width=\linewidth]
{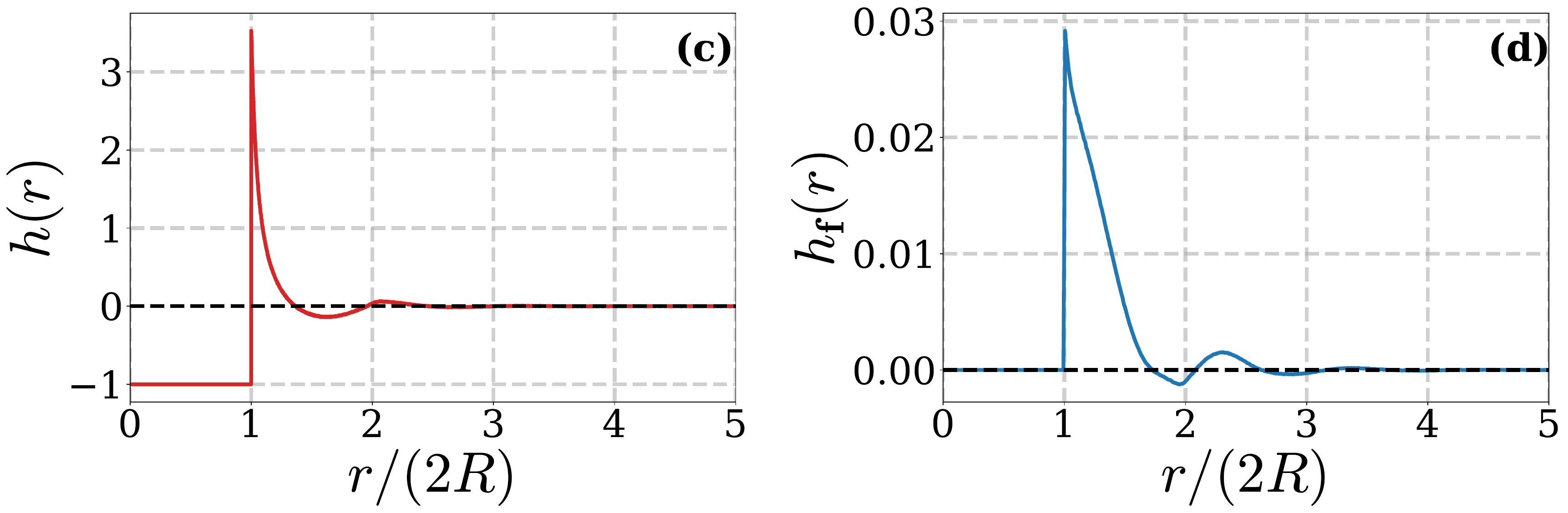}
\caption{Correlation functions for the 3D equilibrium hard-sphere fluid
at $\phi=0.49$, the sampled fluid state closest to freezing (a), (b) and 3D RSA sphere packings at $\phi=0.37$, near saturation (c), (d).
Panels (a) and (c) show $h(r)$, and panels (b) and (d) show $h_{\vf}(r)$,
defined in Eqs.~\eqref{eq:hr} and \eqref{eq:hf}, respectively. The
orientational weight $\vf_j=\{Q_{6m}(j)\}_{m=-6}^{6}$ is constructed using periodic Delaunay neighbors and defined in
Eq.~\eqref{eq:f3D}. Separations are expressed as $r/(2R)$, where $R$
is the sphere radius. In both systems, the correlations decay well within
the size of the simulation boxes.}
\label{fig:correlations_3D_liquid_RSA}
\end{figure}

\begin{figure}[!htb]
\centering
\includegraphics[width=\linewidth]
{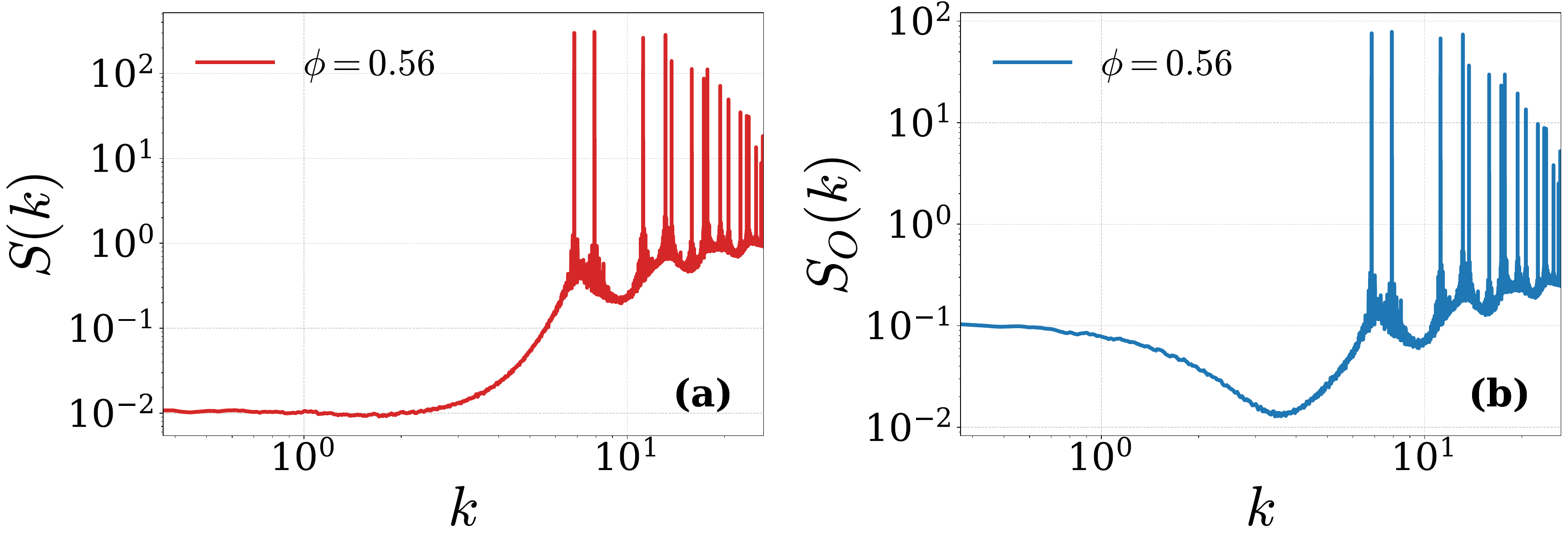}
\caption{Radially averaged structure factors for the 3D equilibrium
hard-sphere fcc crystal at $\phi=0.56$: (a) the translational structure
factor $S(k)$ and (b) the bond-orientational structure factor $S_O(k)$.
The orientational weight
$\vf_j=\{Q_{6m}(j)\}_{m=-6}^{6}$ is constructed using 12 periodic nearest neighbors and defined in Eq.~\eqref{eq:f3D}.}
\label{fig:structure_factors_3D_crystal_phi056}
\end{figure}

In the following, we show representative correlation functions and
structure factors for the model systems described in
Sec.~\ref{sec:models}. Figure~\ref{fig:correlations_2D_liquid_RSA}
shows the correlations for the 2D equilibrium fluid and RSA systems,
while Fig.~\ref{fig:structure_factors_2D_crystal_phi075} shows the
structure factors for the 2D crystal. The 2D disordered SHU structure
factors are shown in
Fig.~\ref{fig:structure_factors_SHU_chi0495}.
Figure~\ref{fig:correlations_3D_liquid_RSA} shows the correlations for
the 3D equilibrium fluid and RSA systems, and
Fig.~\ref{fig:structure_factors_3D_crystal_phi056} shows the structure
factors for the 3D crystal. For both crystal systems, the small-$k$ behavior of
$S(k)$ is consistent with the free-volume-theory prediction~\cite{Wang2024Aug,Torquato2026May}.

\section{Origin and validation of the Poisson reference value}
\label{app:poisson_reference}

\begin{figure}[!htb]
    \centering
    \includegraphics[width=\linewidth]
    {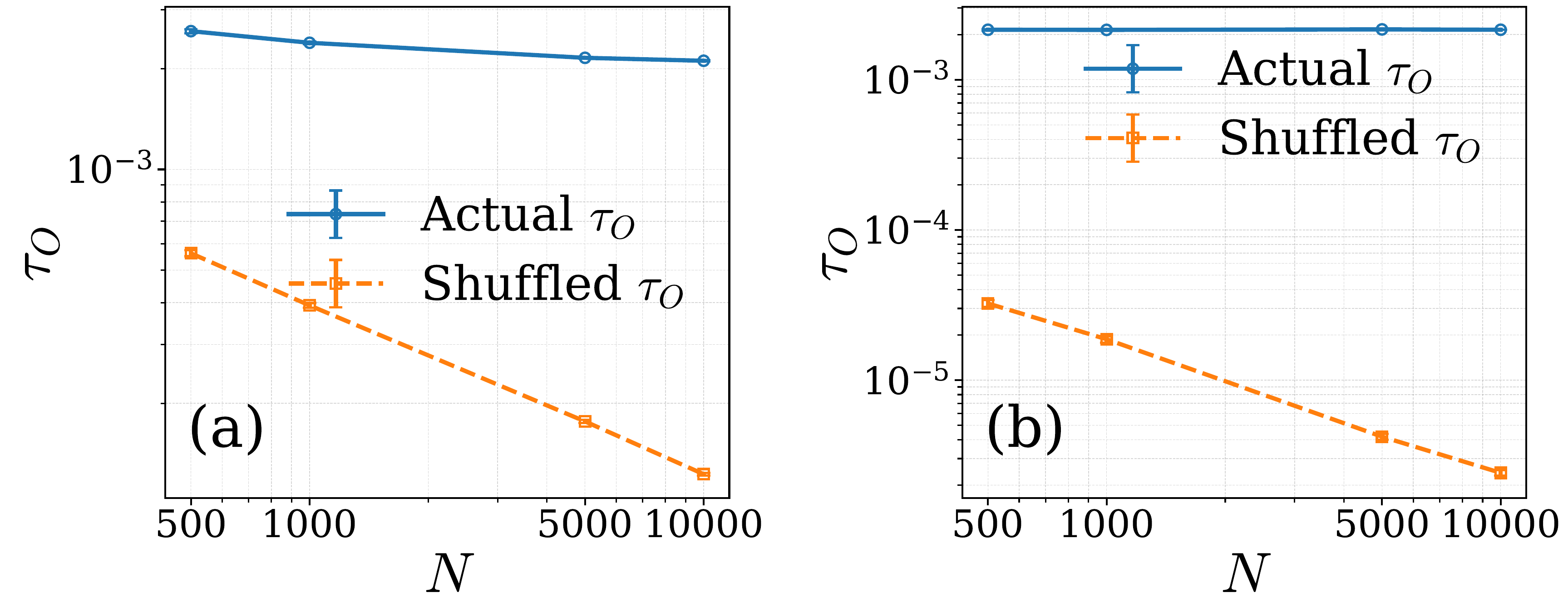}
    \caption{System-size dependence of $\tau_O$ for Poisson point
    configurations in (a) two dimensions and (b) three dimensions, using the
    original Delaunay-derived orientational marks and the same marks
    randomly shuffled among the particle positions. The original values
    approach finite constants, whereas the shuffled values decrease
    toward zero approximately as power laws in $N$.}
    \label{fig:poisson_shuffle}
\end{figure}

For a homogeneous Poisson point process, $g_2(r)=1$ and hence
$h(r)=0$, so that $\tau_T=0$. The orientational marks used to construct
$\tau_O$, however, are not independent marks assigned independently of
the particle positions. They are determined from the local periodic
Delaunay geometry. Nearby particles can therefore have correlated marks
because their local neighbor environments overlap and may contain common
bonds or common neighbors. Thus, although the particle centers form a
Poisson point process and the mean orientational mark vanishes by
isotropy, the weighted total correlation function $h_{\vf}(r)$ need not
vanish at short distances. The resulting nonzero $\tau_O^{\rm P}$
[Eq.~\eqref{eq:tauO_iso_real}] is therefore a
reference value induced by the geometrical construction of the marks,
rather than evidence of macroscopic bond-orientational order.

To separate this geometry-induced contribution from the finite-sampling
background, we randomly shuffle the calculated orientational marks among
the same particle positions. This procedure leaves the point
configuration and the one-particle distribution of the marks unchanged,
but removes the association between each mark and the local geometry
from which it was calculated. Consequently, any persistent difference
between the original and shuffled results isolates the contribution
arising from correlations among the geometry-derived marks.

Figure~\ref{fig:poisson_shuffle} shows the system-size dependence of the
original and shuffled values in two and three dimensions. In both cases,
the original Poisson value approaches a finite constant as the number of
particles increases, whereas the shuffled value decreases toward zero
approximately as a power law in $N$. This behavior confirms that the
nonzero Poisson reference value of $\tau_O$ survives in the
thermodynamic limit and originates from short-range correlations among
the Delaunay-derived orientational marks, while the shuffled contribution
is a finite-sampling effect.

\section{Comparison of $\tau$ metrics for the 3D equilibrium crystal
branch using periodic Delaunay neighbors}
\label{app:3D_crystal_delaunay}

\begin{figure}[!h]
\centering
\includegraphics[width=\linewidth]
{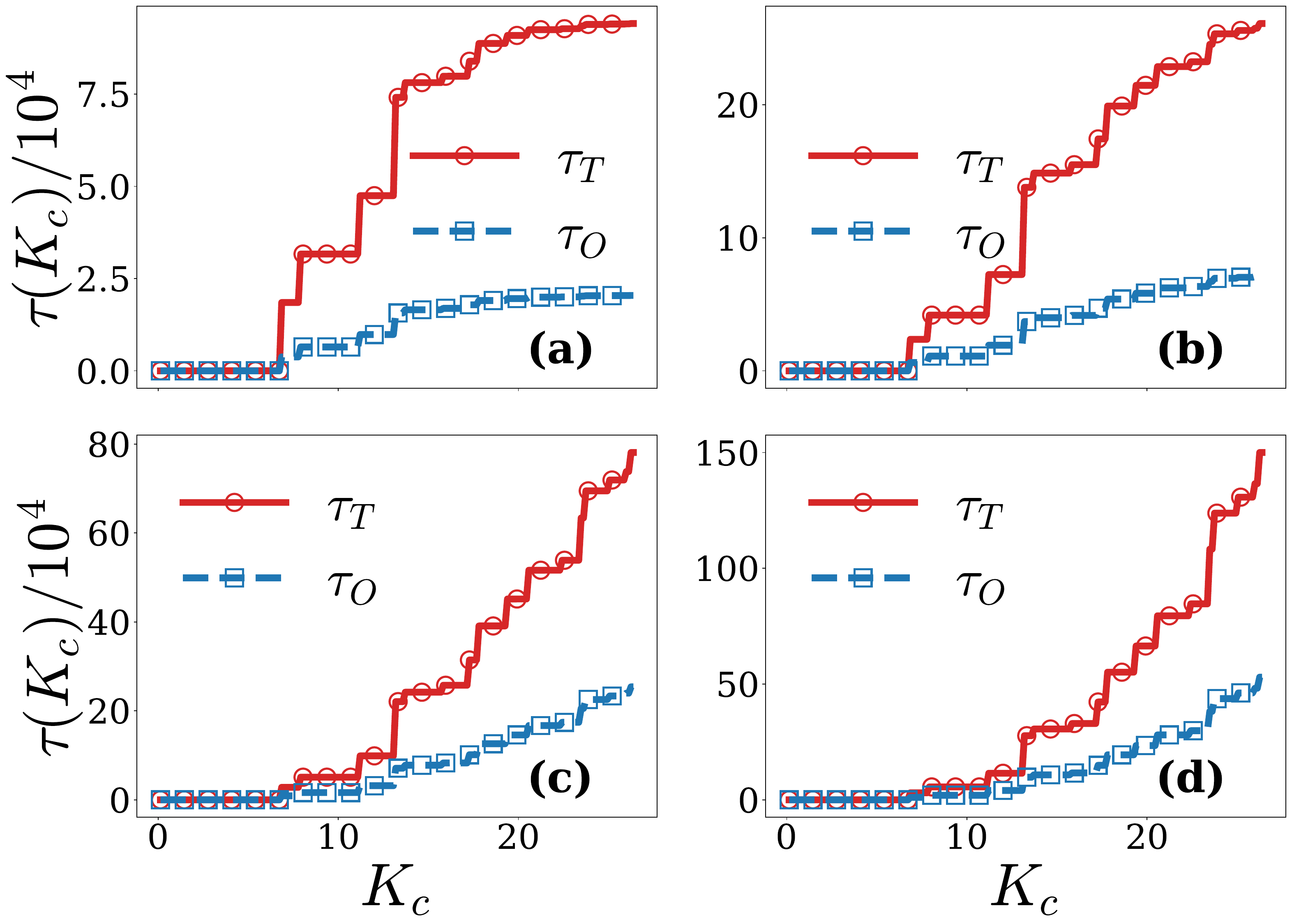}
\includegraphics[width=\linewidth]
{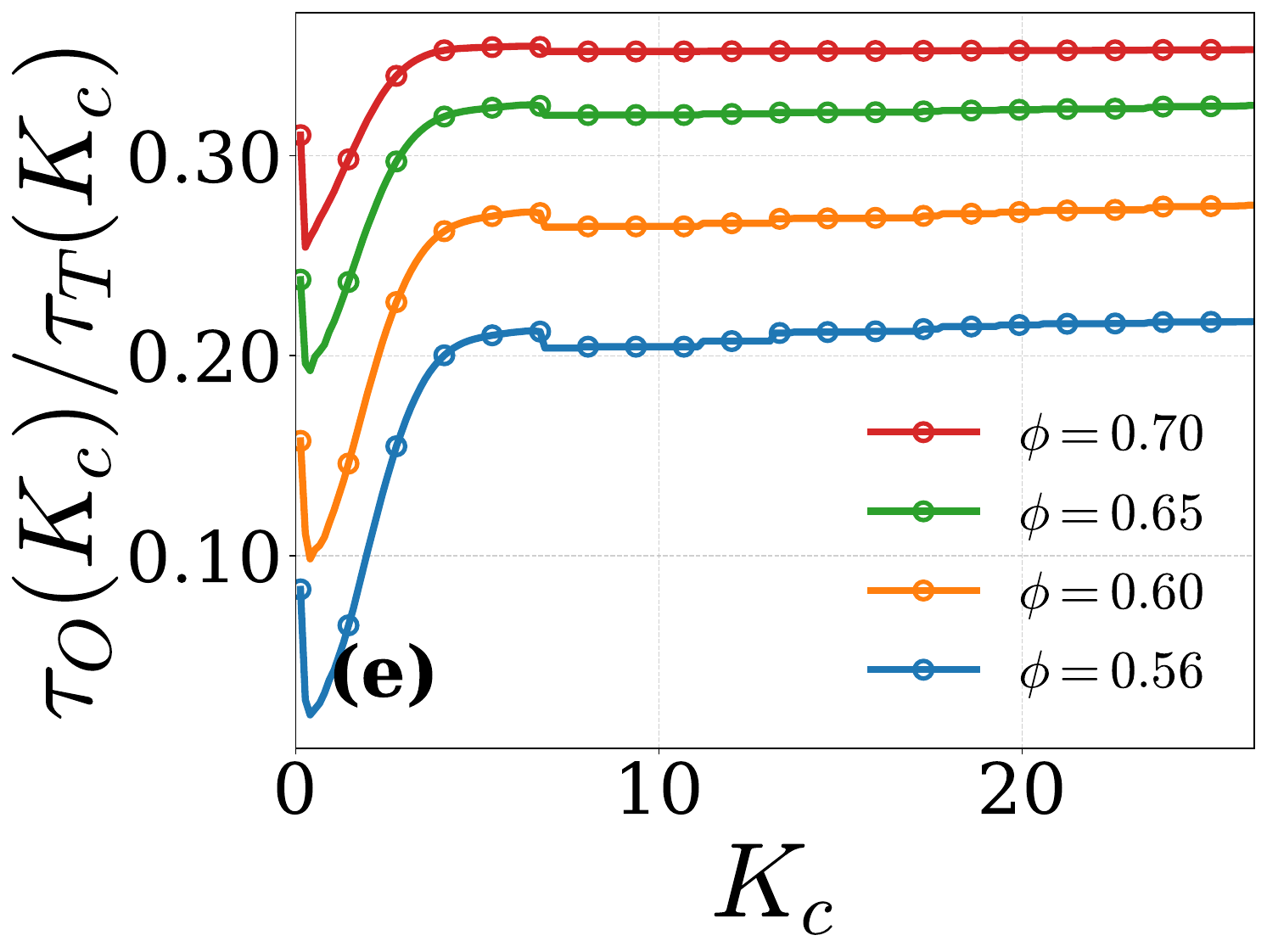}
\caption{Corresponding results for the 3D fcc crystal branch using
periodic Delaunay neighbors as a diagnostic comparison at fixed system
size $L$. The additional neighbors included by the Delaunay
construction suppress $\tau_O$ and consequently reduce
$\tau_O/\tau_T$, while leaving $\tau_T$ unchanged.}
\label{fig:eq_crystal_3D_delaunay_check}
\end{figure}

Periodic Delaunay neighbors provide a diagnostic test of the neighbor
definition. As in the primary crystal calculation,
$\tau_T(K_c)$ and $\tau_O(K_c)$ are evaluated at fixed system size $L$
using the same common radial cutoff $K_c$. Because $\tau_T$ depends
only on particle positions, it is unchanged, whereas $\tau_O$ depends
on the bonds used to construct the local $Q_6$ weights. The Delaunay
construction includes neighbors beyond the 12-neighbor first
coordination shell, reducing the local $Q_6$ amplitudes and suppressing
$\tau_O$. At $\phi=0.70$, the ratio approaches approximately $0.35$
over the higher-$K_c$ range shown
[Fig.~\ref{fig:eq_crystal_3D_delaunay_check}]. We therefore use the
12-neighbor definition, which isolates the first coordination shell
of the fcc lattice, for the primary results presented in
Sec.~\ref{subsubsec:results_crystal_branches} and
retain the Delaunay calculation solely as a diagnostic of the
sensitivity of $\tau_O$ to the neighbor definition.

\end{appendix}

\bibliography{paper}

\end{document}